\documentclass[fleqn,usenatbib]{mnras}

\usepackage{newtxtext,newtxmath}

\usepackage[T1]{fontenc}

\DeclareRobustCommand{\VAN}[3]{#2}
\let\VANthebibliography\thebibliography
\def\thebibliography{\DeclareRobustCommand{\VAN}[3]{##3}\VANthebibliography}

\def \update {}
\def \secupdate {}

\usepackage{graphicx}	
\usepackage{amsmath}	

\usepackage{subfigure}
\graphicspath{{./}{picture/}}
\usepackage{booktabs}

\title[Galaxy Rotation and the Motion of Neighbours]{The SAMI Galaxy Survey: The Relationship between Galaxy Rotation and the Motion of Neighbours}

\author[Y. Mai et al.]{Yifan Mai,$^{1,2}$\thanks{E-mail: yifan.mai@sydney.edu.au}
Sam P. Vaughan,$^{1,2}$
Scott M. Croom,$^{1,2}$
Jesse van de Sande,$^{1,2}$
Stefania Barsanti,$^{2,3}$
\newauthor
Joss Bland-Hawthorn,$^{1,2}$
Sarah Brough,$^{2,4}$
Julia J. Bryant,$^{1,2,5}$
Matthew Colless,$^{2,3}$
Michael Goodwin,$^{6}$
\newauthor
Brent Groves,$^{3}$
Iraklis S. Konstantopoulos,$^{7}$
Jon S. Lawrence,$^{6}$
Nuria P. F. Lorente,$^{6}$
and Samuel N. Richards$^{8}$
\\
$^{1}$Sydney Institute for Astronomy (SIfA), School of Physics, The University of Sydney, NSW, 2006, Australia\\
$^{2}$ARC Centre of Excellence for All Sky Astrophysics in 3 Dimensions (ASTRO 3D)\\
$^{3}$Research School of Astronomy \& Astrophysics, Australian National University, Mount Stromlo Observatory, Cotter Road, Weston Creek, ACT 2611 Australia\\
$^{4}$School of Physics, University of New South Wales, NSW 2052, Australia\\
$^{5}$Australian Astronomical Optics, AAO-USydney, School of Physics, University of Sydney, NSW 2006, Australia\\
$^{6}$Australian Astronomical Optics - Macquarie, Macquarie University, NSW 2109, Australia\\
$^{7}$Independent scholar, Wellington, New Zealand\\
$^{8}$SOFIA Science Center, USRA, NASA Ames Research Center, Building N232, M/S 232-12, P.O. Box 1, Moffett Field, CA 94035-0001, USA\\
}

\date{Accepted 2022 June 22. Received 2022 June 20; in original form 2021 September 23}

\pubyear{2021}

\begin{document}
\label{firstpage}
\pagerange{\pageref{firstpage}--\pageref{lastpage}}
\maketitle

\begin{abstract}
Using data from the SAMI Galaxy Survey, we investigate the correlation between the {\update projected} stellar kinematic spin vector of {\update 1397} SAMI galaxies and the line-of-sight motion of their neighbouring galaxies. We calculate the luminosity-weighted mean velocity difference between SAMI galaxies and their neighbours {\update in the direction perpendicular to the SAMI galaxies' angular momentum axes}. The luminosity-weighted mean velocity {\update offsets between SAMI and neighbours}, which indicates the signal of coherence between the rotation of the SAMI galaxies and the motion of neighbours, is $9.0\pm5.4$ km s$^{-1}$ (1.7$\sigma$) for neighbours within 1 Mpc. In a large-scale analysis, we find that {\update the average velocity offsets increase for neighbours out to 2 Mpc.} However, the velocities are consistent with zero or negative for neighbours outside 3 Mpc. The negative signals for neighbours at distance around 10 Mpc are also significant at $\sim 2\sigma$ level, {\update which indicate that the positive signals within 2 Mpc might come from the variance of large-scale structure.} We also calculate average velocities of different subsamples, including galaxies in different regions of the sky, galaxies with different stellar masses, galaxy type, $\lambda_{Re}$ and inclination. Although low-mass, high-mass, early-type and low-spin galaxies subsamples show 2--3$\sigma$ signal of coherence for the neighbours within 2 Mpc, the results for different inclination subsamples and large-scale results suggest that the $\sim 2 \sigma$ signals might result from coincidental scatter or variance of large-scale structure. Overall, the modest evidence of coherence signals for neighbouring galaxies within 2 Mpc needs to be confirmed by larger samples of observations and simulation studies.

\end{abstract}




\section{Introduction}

The environment in which a galaxy resides plays an important role in its formation and evolution. For example, it is well known that star formation rate is dependent on local galaxy density at low-redshift \citep[e.g.][]{Hashimoto:1998,Kauffmann:2004,Lewis:2002}. {\update Likewise, there is evidence that a galaxy's morphology is also dependent on environment. The morphology-density relation, first presented by \citet{Dressler:1980}, shows that the fraction of elliptical and S0 galaxies increases with increasing density while the fraction of spiral galaxies decreases.} {\update \citet[][]{Cappellari:2011} found the existence of a kinematic morphology-density relationship, which showed that slow-rotator galaxies are more likely to be found in high density environments. \citet[][]{Sande:2021} also found that environmental density correlates with the fraction of slow rotators after accounting for the effect of stellar mass \citep[but see also][]{Brough:2017,Greene:2017,Veale:2017}.} 

A key question that relates to morphology is how galaxies acquire their angular momentum. Tidal torque theories state that protogalaxies acquire angular momentum through torquing moments exerted by nearby large-scale structure. The angular momentum of protogalaxies is conserved during collapse \citep{Doroshkevich:1970,White:1984,Porciani:2002,Schafer:2009}. However, tidal torque theories are not enough to give a full picture of the build up of angular momentum of present-day galaxies. The spin of a galaxy may deviate from the original direction due to non-linear effects such as mergers \citep{Welker:2014}. 

A different and novel investigation into the relationship between a galaxy and its local environment was conducted by \cite{Welker:2020}. They used data from the SAMI Galaxy Survey to find how a galaxy's spin axis is related to the nearest galaxy filament. \citet{Welker:2020} found that the spin of low-mass galaxies tended to be aligned with the nearest filament while the spin of high-mass galaxies was more likely to be orthogonal to their nearest filament. Their results are consistent with other observational studies \citep{Tempel:2013,Kraljic:2021} and numerical simulation results \citep{Codis:2012,Dubois:2014,Wang:2018,Ganeshaiah:2019,Kraljic:2020}. {\update Most recently, Barsanti et al. (submitted), using the SAMI Galaxy Survey, find that the mass of the bulge is the primary parameter of correlation with spin-filament alignments.}

\defcitealias{Lee:2019}{L19a}
\defcitealias{Lee:2019(b)}{L19b}

\citet[][hereafter L19a]{Lee:2019} took a different approach, examining whether the rotation of a galaxy is correlated with the motion of its neighbours. {\update Their results can be considered as observational evidence that the direction of galaxy rotation is influenced by the interaction of the neighbouring galaxies through adding new angular momentum during flybys and/or mergers.} They estimated the angular momentum vectors of 445 galaxies from the Calar Alto Legacy Intergral Field Area survey \citep[CALIFA;][]{sanchez:2012} whilst also measuring the relative motion of their neighbouring galaxies by comparing the galaxy redshifts. Then, they built a composite map of the velocity distribution of each galaxy's neighbours and calculated the luminosity-weighted mean velocity as a function of radius from the central galaxy. They found that coherence between the rotations of CALIFA galaxies and the motion of neighbouring galaxies is significant for the neighbours up to 800 kpc in projected distance on the sky. They also found hints that the coherence signal between galaxies and their neighbours was stronger when they only studied the outskirts of each galaxy, rather than when they studied the central regions, and that the spin of low-mass galaxies tended to be more aligned with their neighbours than the rotation of high-mass galaxies. 

\citet[][hereafter L19b]{Lee:2019(b)} investigated this coherence on larger scales. They extended the scope of neighbour galaxies to 15 Mpc, and found that this coherence exists out to several Mpc scales, especially between the central rotation of CALIFA galaxies and the neighbouring galaxies with red optical colours. {\update This result might indicate a correlation between the long-term motion of the large-scale structure and the direction of a galaxy's rotation.} \citetalias{Lee:2019(b)} found that the absolute-luminosity-weighted velocity for central rotation (30.6 $\pm$ 10.9 km s$^{-1}$) is slightly larger than that for outskirt rotation (25.9 $\pm$ 11.2 km s$^{-1}$) when measured using the neighbours at distance 1 Mpc < $D$ < 6 Mpc. On the contrary, \citetalias{Lee:2019} found that the coherence signal is stronger for the outskirt rotation of CALIFA galaxies when considering the neighbours within 1 Mpc. \citetalias{Lee:2019(b)} also analysed the subsamples of CALIFA galaxies and found that the high Sérsic index galaxies (Sérsic index $n$ > 2) or internally-well-aligned galaxies (position
angle difference between the central and outskirt
angular momentum vectors is smaller than 5.0°) have slightly stronger coherence.

The \citetalias{Lee:2019} and \citetalias{Lee:2019(b)} results are intriguing, but the relatively small sample size of CALIFA galaxies hampers the ability to draw any firm conclusions about the size of the coherence signal. In this study, we investigate the existence of coherence between the angular momentum vectors of galaxies and the motion of their neighbours, carrying out analysis similar to \citetalias{Lee:2019} and \citetalias{Lee:2019(b)}, using a sample that is three times as large from the SAMI Galaxy Survey. The plan of the paper is as follows: in Section \ref{sec:Data} and Section \ref{sec:methods}, we describe the data and methods of this study; in Section \ref{sec:results} and Section \ref{sec:discussion}, we analyse the data and discuss our results; in Section \ref{sec:Sum}, we present our conclusions. Throughout this paper, we adopt the concordance cosmology: ($\Omega_{\Lambda}$,$\Omega_{m}$,$h$) = (0.7, 0.3, 0.7).

\section{Data} \label{sec:Data}
\begin{figure*}
\centering
\includegraphics[width=0.9\textwidth]{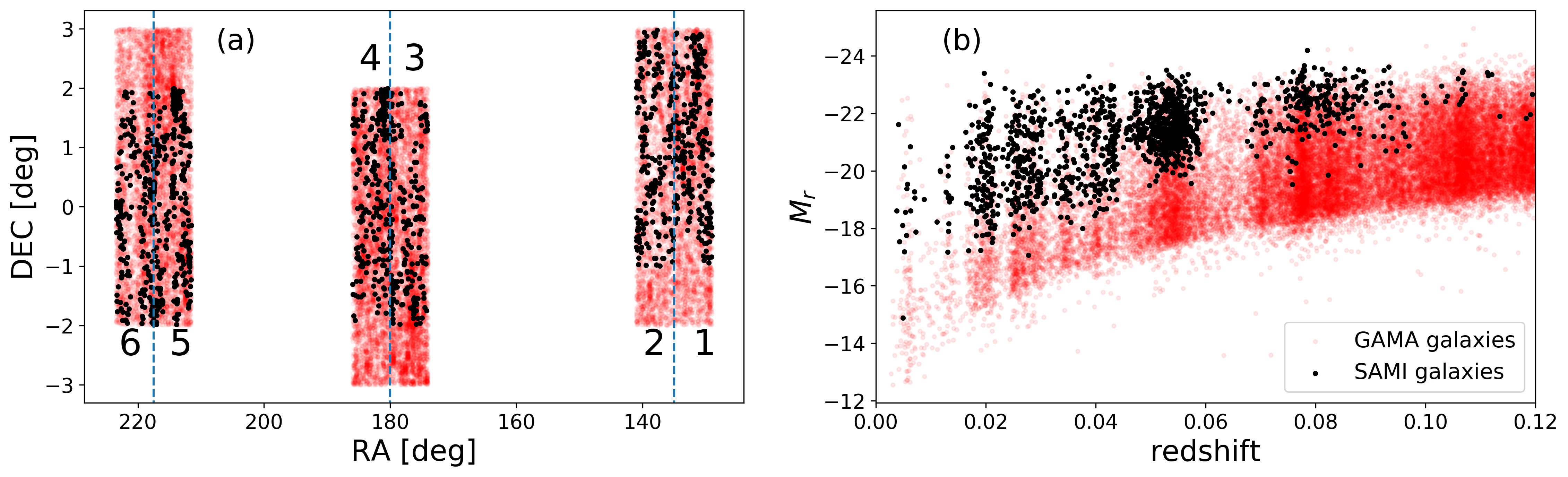}
\caption{(a) Red points indicate the spatial distribution of GAMA galaxies, and black points show the spatial distribution of the SAMI galaxies. We divided
the SAMI galaxies into 6 parts (as indicated by black numbers) in order to investigate the variance of large scale structure between different regions; (b) $r$-band absolute magnitude versus redshift diagram of SAMI and GAMA galaxies. \label{fig:SpatialDist&z-mag}}
\end{figure*}

We use data of galaxies observed by the Sydney-AAO (Australian Astronomical Observatory) Multi-object Integral field spectrograph \citep[SAMI;][]{Croom:2012} as part of the SAMI Galaxy Survey \citep{Bryant:2015}. SAMI has collected spatially resolved spectroscopy for 3068 galaxies between a redshift of 0.004 and 0.115, with wavelength coverage between 3700--5700 \AA\ in the blue and 6300--7400 \AA\ in the red. The SAMI integral field units are hexabundles \citep{bryant:2014,Bland-Hawthorn:2011} with radius 7.5 arcsec, consisting of 61 individual fibres, whose radii are 0.8 arcsec. SAMI fibres are fed to the double-beam AAOmega spectrograph \citep{Sharp:2006}. 40 percent of the galaxies are observed to more than twice their effective radius ($R_e$), and the $R_e$ of 17 percent galaxies are larger than SAMI bundle. We use data from the third public SAMI data release \citep[DR3;][]{croom:2021}, which presents fully reduced data-cubes as well as catalogues of galaxy properties and maps of spatially resolved kinematics \citep{Sharp:2015,Bryant:2015,Owers:2017}. The SAMI galaxies are distributed in three 4 $\times$ 12 deg$^{2}$ fields, as shown in Figure \ref{fig:SpatialDist&z-mag}(a). Figure \ref{fig:SpatialDist&z-mag}(b) shows the $r$-band absolute magnitude versus redshift diagram of the SAMI galaxies {\update we use in this study} (black points) and the neighbouring galaxies (red points). {\update The SAMI survey includes four volume limited samples which consist of a stepped series of stellar mass limits \citep[see][]{Bryant:2015}.} We do not include SAMI cluster galaxies in this sample.

This work defines the “neighbour galaxies” as the galaxies that have line-of-sight velocity differences {\update with respect to the given SAMI galaxies} less than 500 km s$^{-1}$ and projected distances smaller than 20 Mpc. The neighbour galaxies database comes from the Galaxy and Mass Assembly (GAMA) survey \citep{Driver:2011}, which is a spectroscopic survey carried out using the AAOmega multi-object spectrograph on the Anglo-Australian Telescope (AAT). The GAMA survey consists of five individual regions. We used the data of galaxies in three equatorial regions called G09, G12 and G15 each of 60 deg$^{2}$ \citep{baldry:2018,Liske:2015}, within which the SAMI Galaxy Survey has been observed.

In this study, the main quantity of interest is the stellar kinematic
position angle (PA) of each galaxy. We use the measurements of PA from \citet{Sande:2017}, and briefly outline the measurement method below. PA represents the angle between the negative velocity peak (with respect to the systematic velocity of the galaxy) of the kinematic map and North. For example, a kinematic PA of 90° represents the negative velocity part of that galaxy aligned to the east. PA is measured using the FIT\_KINEMATIC\_PA code, based on the method presented in Appendix C of \citet{Krajnovic:2006}. PAs are obtained from the two-dimensional stellar velocity kinematic maps using spaxels that have uncertainties in stellar velocity less than 30 km s$^{-1}$. Other than the limit on uncertainty, we do not place any further constraints on which spaxels are used, in contrast with \citetalias{Lee:2019}. They measured the PA of CALIFA galaxies from the central ($R\leq R_e$) and outskirt regions ($R_e<R\leq 2R_e$) respectively. {\update Because the bundles used in SAMI survey only cover up to 1.5 $R_e$ for most of SAMI targets, we are not able to perform the measurement for $R_e<R\leq 2R_e$. Therefore, we use all spaxels to measure PA. Our results are comparable to the "$R\leq R_e$" subsample on \citetalias{Lee:2019}, where they only use spaxels within $R_e$ to measure PA.}

\begin{figure}
\centering
\includegraphics[width=0.45\textwidth]{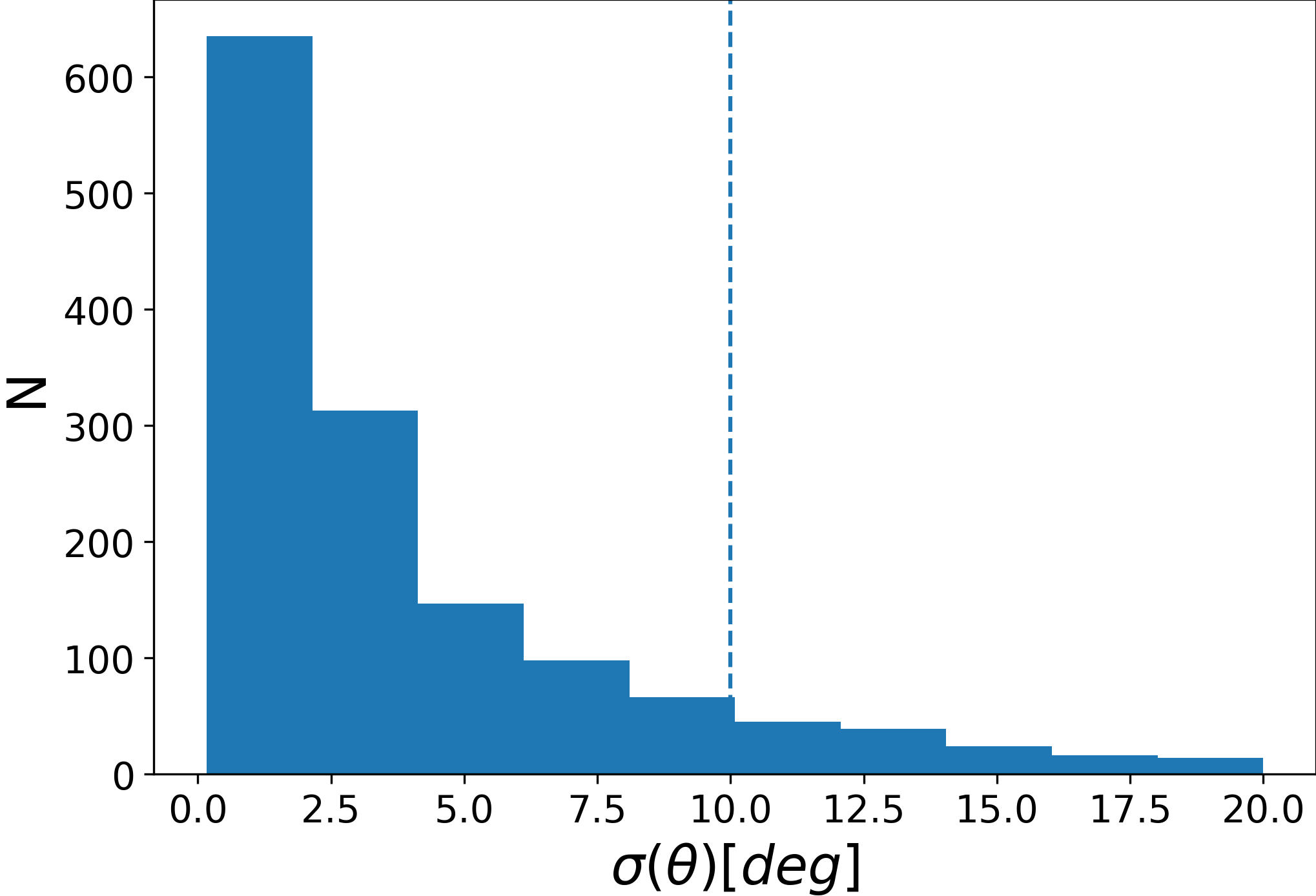}
\caption{Uncertainty distribution of the position angles of the SAMI galaxies. The vertical line denotes 10° uncertainty.}
\label{fig:uncertDistr}
\end{figure}

The SAMI DR3 includes 1559 galaxies with PA data in the GAMA regions. In order to decrease the negative effects of position angle uncertainties, we only use data where the error on PA is smaller than 10°, resulting in a sample of 1257 galaxies. In order to compare our results with \citetalias{Lee:2019}, we also process the data using an error limit of $<20$°. This sample contains 1397 galaxies. Figure \ref{fig:uncertDistr} demonstrates the uncertainty distribution of the position angles of the SAMI galaxies that match criteria. {\update Though the target selection of SAMI survey aims to cover a broad range in stellar mass, galaxies with poor measurement of stellar kinematic position angle tend to be faint.}

The stellar mass of SAMI galaxies are measured using a method described in \citet{Bryant:2015}. The morphologies of SAMI galaxies have been visually classified by SAMI team members \citep{Cortese:2016}. The measurements of $\lambda_{Re}$ are conducted by \citet{Sande:2017} with aperture correction \citep{sande:2017MNRAS} and seeing correction from \citet{Harborne:2020} (optimised for SAMI in \citealt{Sande:2021SeeingCorr}). The measurement of photometric ellipticity \citep{DEugenio:2021} is based on $r$-band SDSS and {\update VLT Survey Telescope (VST) \citep{Shanks:2015}} images using Multi Gaussian Expansion \citep[MGE,][]{Cappellari:2002}. We obtain an $r$-band axis ratio $(b/a)$ from ellipticity and determine the inclination as follows \citep{Hubble:1926}:

\begin{equation}
    \cos \left ( i \right ) = \sqrt{\frac{\left ( b/a \right )^{2}-\left ( b/a \right ) _{\mathrm{min} }^{2}   }{1-\left ( b/a \right ) _{\mathrm{min} }^{2} } } ,
\end{equation}

\noindent with $(b/a)_{\mathrm{min}}=0.15$ (following \citealt{Sargent:2010} and \citealt{Leslie:2018} and based on \citealt{Guthrie:1992} and \citealt{Yuan:2004}).

In order to investigate the relative motion between the SAMI galaxies and their neighbours, we calculate the line-of-sight velocity offset ($\Delta\ v$ ) for each neighbouring galaxy as 
\begin{equation}
    \Delta v=\frac{z_{\textup{nei}}-z_{\textup{SAMI}}}{1+z_{\textup{SAMI}}}\cdot c,
\end{equation}
 where $z_{\textup{nei}}$ is the heliocentric redshift of neighbour galaxies around SAMI galaxies, $z_{\textup{SAMI}}$ is the heliocentric redshift of SAMI galaxies and $c$ is the speed of light. Both redshifts of neighbour galaxies and SAMI galaxies are from the GAMA catalogue, as galaxies observed by SAMI are drawn from the GAMA survey.

\section{Methods} \label{sec:methods}
\subsection{Measurements of average velocities of neighbouring galaxies}

\begin{figure*}
\centering
\includegraphics[width=0.9\textwidth]{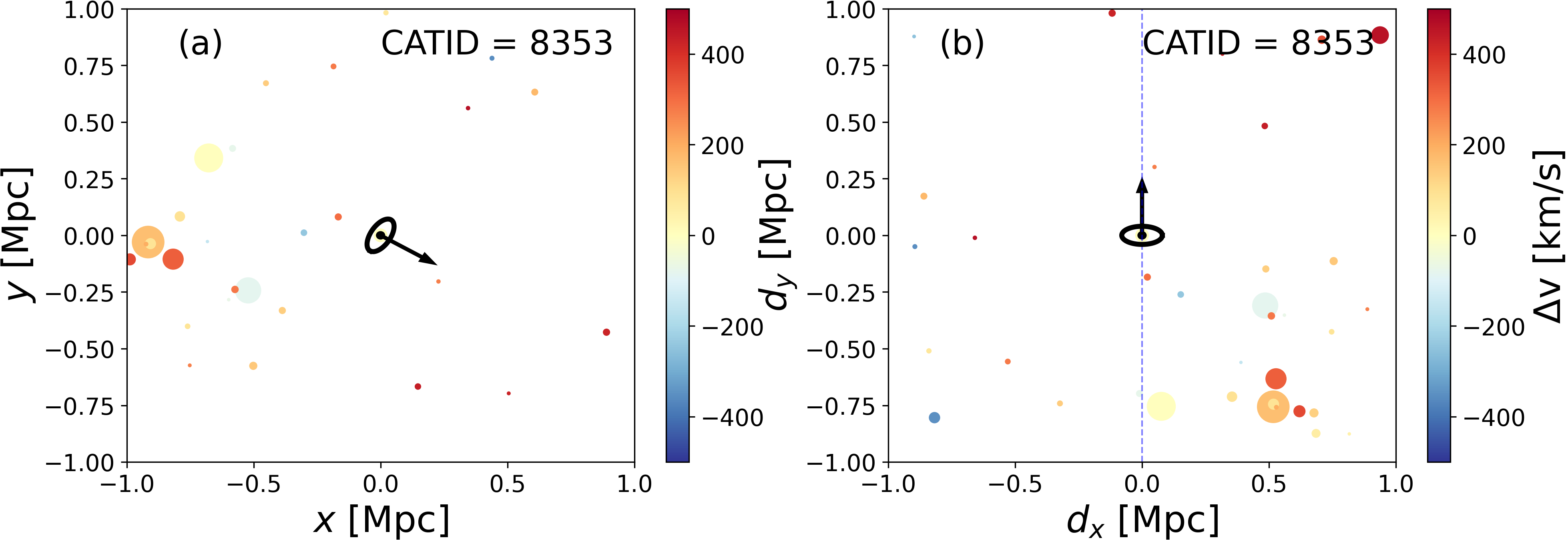}
\caption{The left panel (a) shows a SAMI galaxy and its neighbour galaxies. In the right panel (b) we have rotated the whole picture to make sure that the spin vector (PA) of SAMI galaxy is upward. The black ellipses in the middle of pictures demonstrate an example of SAMI galaxy and the black arrows show its spin vector. The neighbour circles illustrate the neighbour galaxies. The size of circles are proportional to the luminosity of galaxies and the color show the line-of-sight velocity of neighbour galaxies relative to the central SAMI galaxy. \label{fig:sampleGalaxy}}
\end{figure*}

\begin{figure*}
\centering
\includegraphics[width=0.5\textwidth]{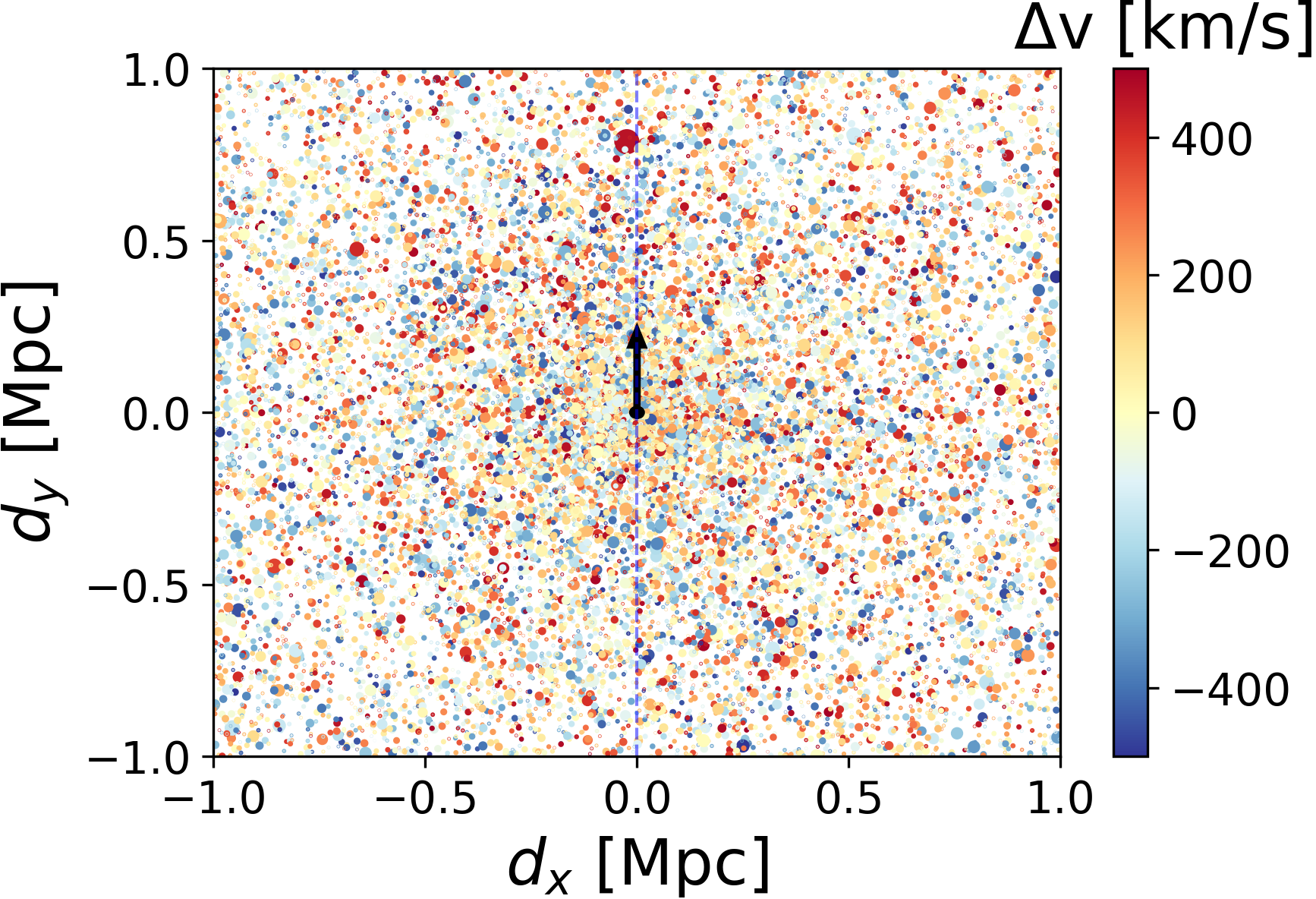}
\caption{We stack 1257 of the maps of neighbours of SAMI galaxies as shown in Figure \ref{fig:sampleGalaxy}(b). The point sizes are proportional to the luminosity of galaxies. The color of points shows the velocity offset between the SAMI galaxies and their neighbours. If there is a strong signal, the right side will be predominantly red, the left side will be predominantly blue. \label{fig:composite1257}}
\end{figure*}

In order to investigate the relationship between galaxies' rotation and the motion of their neighbours, we compare the direction of the projected angular momenta of SAMI galaxies with the line-of-sight velocity offset of their neighbour galaxies. This method is based on that of \citetalias{Lee:2019}. We rotate all the galaxies in a 20 Mpc by 20 Mpc box around each SAMI galaxy so that the SAMI galaxy has its angular momentum axis pointing vertically (North). This means that rotation towards the observer (negative velocity) is to the East and rotation away from the observer (positive velocity) is to the West. Figure \ref{fig:sampleGalaxy}(a) shows {\update an example of a SAMI galaxy} and the relative motion of its neighbour galaxies, while Figure \ref{fig:sampleGalaxy}(b) shows the map after rotation. The arrow indicates the direction of the projected angular momentum of the SAMI galaxy. Then, we stacked these maps for all 1257 SAMI galaxies in our sample to investigate the overall trend of neighbour galaxies motion, as shown in Figure \ref{fig:composite1257}. If a correlation between the angular momentum vector of galaxies and the motion of their neighbours is sufficiently strong, the left side of Figure \ref{fig:composite1257} would be predominantly blue and the right side would be predominantly red.

\begin{figure*}
\centering
\includegraphics[width=0.9\textwidth]{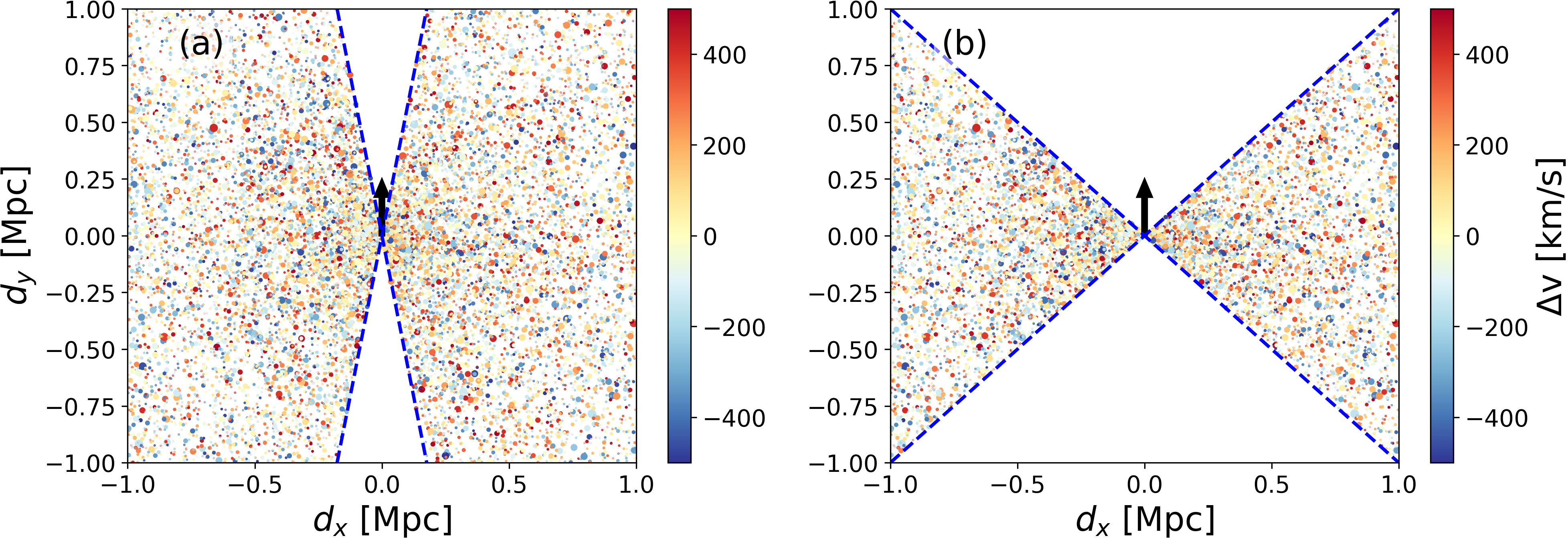}
\caption{Two methods of choosing neighbouring galaxies. (a) X-cut-10°: preserve the data of neighbour galaxies located between 10° and 170° or between 190° and 350° from north. (b) X-cut-45°: preserve the data of neighbour galaxies located between 45° and 135° or between 225° and 315° from north.  \label{fig:compositeX-cut}}
\end{figure*}

We then select which neighbouring galaxies would form part of our quantitative analysis in two different ways:

\begin{enumerate}
\item X-cut-10°: we form this subsample by removing neighbouring galaxies which are located $\pm$ 10° from the North and South directions, as shown in Figure \ref{fig:compositeX-cut}(a).
\item X-cut-45°: In order to compare our results with L19, we preserved the galaxies with PA uncertainty smaller than 20° and removed neighbouring galaxies which are located $\pm$ 45° from the North and South directions, as shown in Figure \ref{fig:compositeX-cut}(b).
\end{enumerate}

The selection (ii) is the same as used by L19, but the PA uncertainty of SAMI data is smaller. The PA uncertainty cuts are 10° in (i) and 20° in (ii), whilst \citetalias{Lee:2019} used 45°.

We then calculate the average line-of-sight velocity of the neighbouring galaxies as a function of projected {\update comoving} distance $d$ for both (i) and (ii), 

{\update
\begin{equation}
    d = \textup{sign}\left ( d_x \right ) \cdot \sqrt{d_{x}^{2}+d_{y}^{2}},
\end{equation}
}

{\update where $d_x$ is projected comoving distance perpendicular to the direction of angular momentum of the given target SAMI galaxy and $d_y$ is projected comoving distance parallel to the direction of angular momentum.}

We use three kinds of luminosity-weighted average to calculate the mean line-of-sight velocity offset of neighbour galaxies. The first method is absolute-luminosity (abs-L) weighting, which averages the velocity of neighbouring galaxies weighted by their $r$-band luminosity. This method does not take into account the luminosity of the SAMI galaxies from which we measured angular momentum vectors. Secondly, we use a relative-luminosity (rel-L) weighting, which weights each neighbour galaxy's line-of-sight velocity by the ratio between the $r$-band luminosity of the neighbouring galaxy and the $r$-band luminosity of the SAMI galaxy. These two kinds of luminosity-weighted methods are the same as that in \citetalias{Lee:2019}, which helps us to compare our results with theirs. Finally, we also calculated an unweighted average, treating each of the neighbouring galaxies equally. This is equivalent to thinking of the galaxies as tracer particles in the overall dark matter distribution. In contrast, the luminosity weighting approach implies that higher luminosity galaxies are associated with higher mass, potentially leading to higher torques of the SAMI galaxies.

Following equations 2, 3 and 4 in \citetalias{Lee:2019}, {\update we calculate the mean velocity offset at ${D}$, where $D$ is the distance for a bin instead of individual galaxies.} This method of finding the average velocity difference of galaxies in a 100 kpc window is used to smooth the profile, which is otherwise dominated by scatter, such that

\begin{equation}
\left \langle \Delta v \right  \rangle^{d100}\left (  {D} \right ) =  
\left\{
    \begin{array}{ll}
    \frac{\sum _{Rd\left ( {D},100 \right )}\Delta v L_{\textup{w}}}{\sum _{Rd\left ( {D},100 \right )} L_{\textup{w}}}&\textup{if}\ {D}>0\\
    0 & \textup{if}\ {D}=0 \\ 
    \frac{\sum _{Ld\left ( {D},100 \right )}\Delta v L_{\textup{w}}}{\sum _{Ld\left ( {D},100 \right )} L_{\textup{w}}}&\textup{if}\ {D}<0.
    \end{array}
\right.
\label{equ:radiusProfile}
\end{equation}

\noindent We take $L_{\textup{w}}$ to be the absolute luminosity of the neighbouring galaxy in the abs-L case, and the ratio of luminosity of the SAMI galaxies and their neighbours in the rel-L case ($L_{\textup{w}}$=$L_{\mathrm{nei}}$/$L_{\mathrm{SAMI}}$). For the equal-weight case, $L_{\textup{w}}$ is identical for all galaxies. The right-side distance range $R_d$ is

\begin{equation}
\begin{split}
&R_d\left ( {D},100 \right )=\\
&\begin{cases}
    {D}-100\ \textup{kpc} < d\leqslant {D} & \textup{if}\ {D}> 100\ \textup{kpc} \\ 
    0< d\leqslant {D} & \textup{if}\  0<{D}\leqslant 100\ \textup{kpc}, 
\end{cases}
\end{split}
\label{equ:Rd}
\end{equation}

\noindent and the left-side distance range $L_d$ is
\begin{equation}
\begin{split}
&L_d\left ( {D},100 \right )=\\
&\begin{cases}
    {D} \leqslant d< {D}+100\ \textup{kpc} & \textup{if}\ {D}< -100\ \textup{kpc} \\ 
    {D}\leqslant d< 0 & \textup{if}\ -100\ \textup{kpc}\leqslant{D}< 0. 
\end{cases}
\end{split}
\label{equ:Ld}
\end{equation}

This is the approach used by \citetalias{Lee:2019}. However, note that when calculated like this, individual points are not independent of each other.

Furthermore, we simplify the radial profile by merging the left and right hand side mean velocities using

\begin{equation}
\begin{split}
&\left \langle \Delta v \right  \rangle_{R-L}^{d100}\left (  {D} \right )=\\
&\frac{\left (  \sum _{R_d\left ( {D},100 \right )}\Delta v L_{\textup{w}}\right )-\left (  \sum _{L_d\left ( {-D},100 \right )}\Delta v L_{\textup{w}}\right )}{\left (  \sum _{R_d\left ( {D},100 \right )} L_{\textup{w}}\right )+\left (  \sum _{L_d\left ( {-D},100 \right )} L_{\textup{w}}\right )}
\end{split}
\label{equ:R-L_radial}
\end{equation}

\noindent where ${D} > 0$. Figuratively speaking, we fold the profiles from the positive and negative $D$ directions together by flipping the sign and combining. If the coherence exists, the right-left-merged mean velocities would be positive. 

We also create cumulative right-left-merged mean velocities according to:

\begin{equation}
\begin{split}
&\left \langle \Delta v \right  \rangle_{R-L}^{cumu}\left (  {D} \right )=\\
&\frac{\left (  \sum _{0<d\leqslant{D}}\Delta v L_{\textup{w}}\right )-\left (  \sum _{-{D}\leqslant d<0}\Delta v L_{\textup{w}}\right )}{\left (  \sum _{0<d\leqslant{D}} L_{\textup{w}}\right )+\left (  \sum _{-{D}\leqslant d<0} L_{\textup{w}}\right )}.
\end{split}
\label{equ:R-L_cumu}
\end{equation}

\noindent $\left \langle \Delta v \right  \rangle_{R-L}^{cumu}$ can be interpreted as coherence signal within given projected distance ${D}$.

\subsection{Uncertainty estimation}

We use two different methods to estimate the uncertainties on our measurements, and hence analyse the significance of any velocity offsets. The first one is a bootstrapping method, which randomly resamples the neighbour galaxies with replacement 1000 times and calculates the standard deviation of the mean velocity of resamples ($\sigma_{BST}$).

In the second method, we randomize the spin-axis of each SAMI galaxy, build a new composite map and obtain the right-left-merged average velocity. Following \citetalias{Lee:2019}, we call these data "RAX", for "random spin-axis". As we have randomized the PAs, we expect the RAX data to have a mean of zero velocity offset. The scatter about this zero gives us another estimate of the uncertainty on the measurement. We repeat this process 100 times. 

\begin{figure*}
\centering
\includegraphics[width=0.9\textwidth]{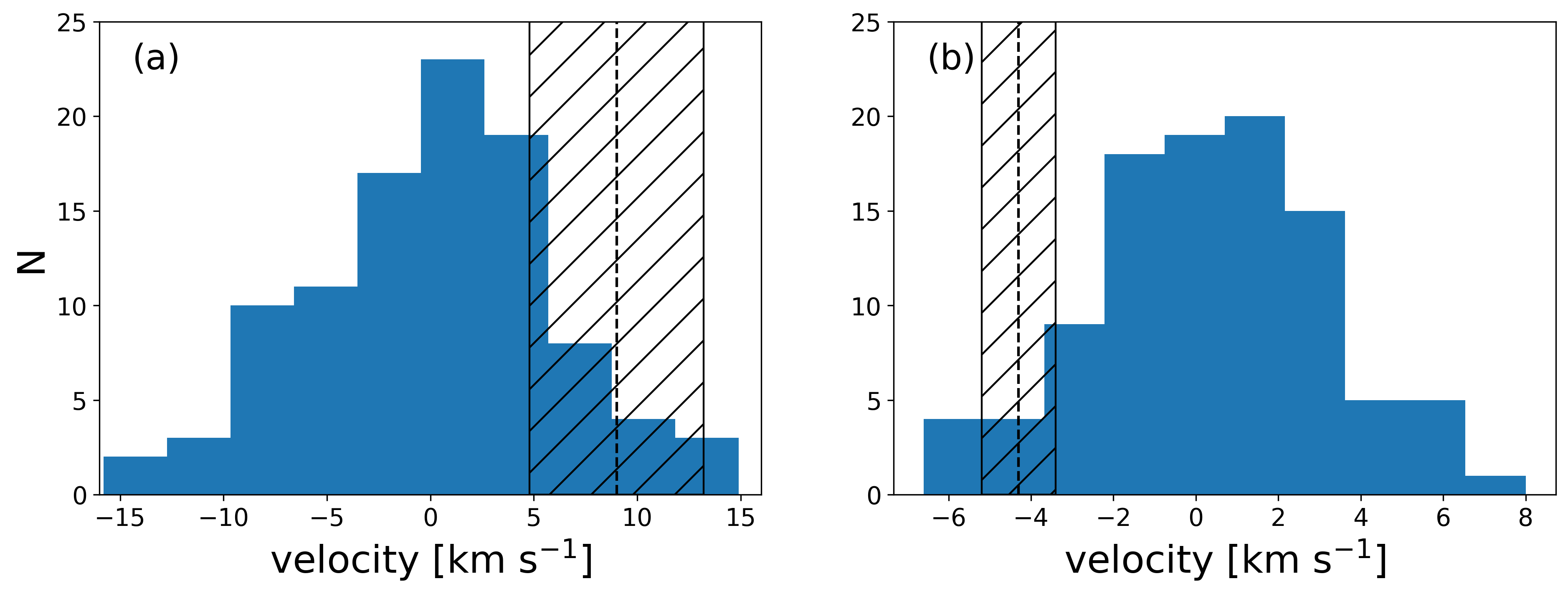}
\caption{(a) The histogram shows the distribution of right-left-merged abs-L weighted average velocity offset between randomized the spin-axis SAMI galaxies and the neighbouring galaxies within 1 Mpc. The standard deviation of the histogram ($\sigma_{RAX}$) is 5.4 km s$^{-1}$. The vertical dashed line marks the average velocity of the real data and the diagonally-shaded region shows its uncertainty using bootstrap estimates. (b) The same as (a), but for the neighbouring galaxies within 20 Mpc. The standard deviation of the histogram is 2.9 km s$^{-1}$. \label{fig:RAX}}
\end{figure*}

Figure \ref{fig:RAX} shows the distribution of average velocities for 100 RAX samples, where the vertical dash lines mark the average velocities of the real data. The diagonal-line-shaded regions are the bootstrap uncertainty estimates. The {\update uncertainties using RAX methods} are 6.0 km s$^{-1}$ for the neighbour galaxies within 1 Mpc and 2.9 km s$^{-1}$ for the neighbour galaxies within 20 Mpc. 
The RAX uncertainty for the abs-L weighting is 1.4 times as large as bootstrap uncertainty (4.2 km s$^{-1}$) for the neighbours within 1 Mpc and 3.2 times as large as bootstrap uncertainty (0.9 km s$^{-1}$) for neighbours within 20 Mpc. Considering the other weighting schemes, we find the RAX uncertainty is 3 times and 5 times greater than bootstrap uncertainties for the rel-L and unweighted cases respectively, which indicates that the bootstrap method may underestimate the variance caused by large-scale structure. When resampling the data for the bootstrap analysis, we select data from mostly the same structures. Bootstrapping, therefore, does not capture the cosmic variance between different patches of sky.

In the following Sections, we will use $\sigma_{RAX}$ as main value of uncertainty, as we consider they give more robust estimation of uncertainty than bootstrapping. 

\section{Results} \label{sec:results}

\subsection{Results for neighbour galaxies within 1 Mpc} \label{sec:results 1Mpc}

\begin{figure*}
\centering
\includegraphics[width=0.9\textwidth]{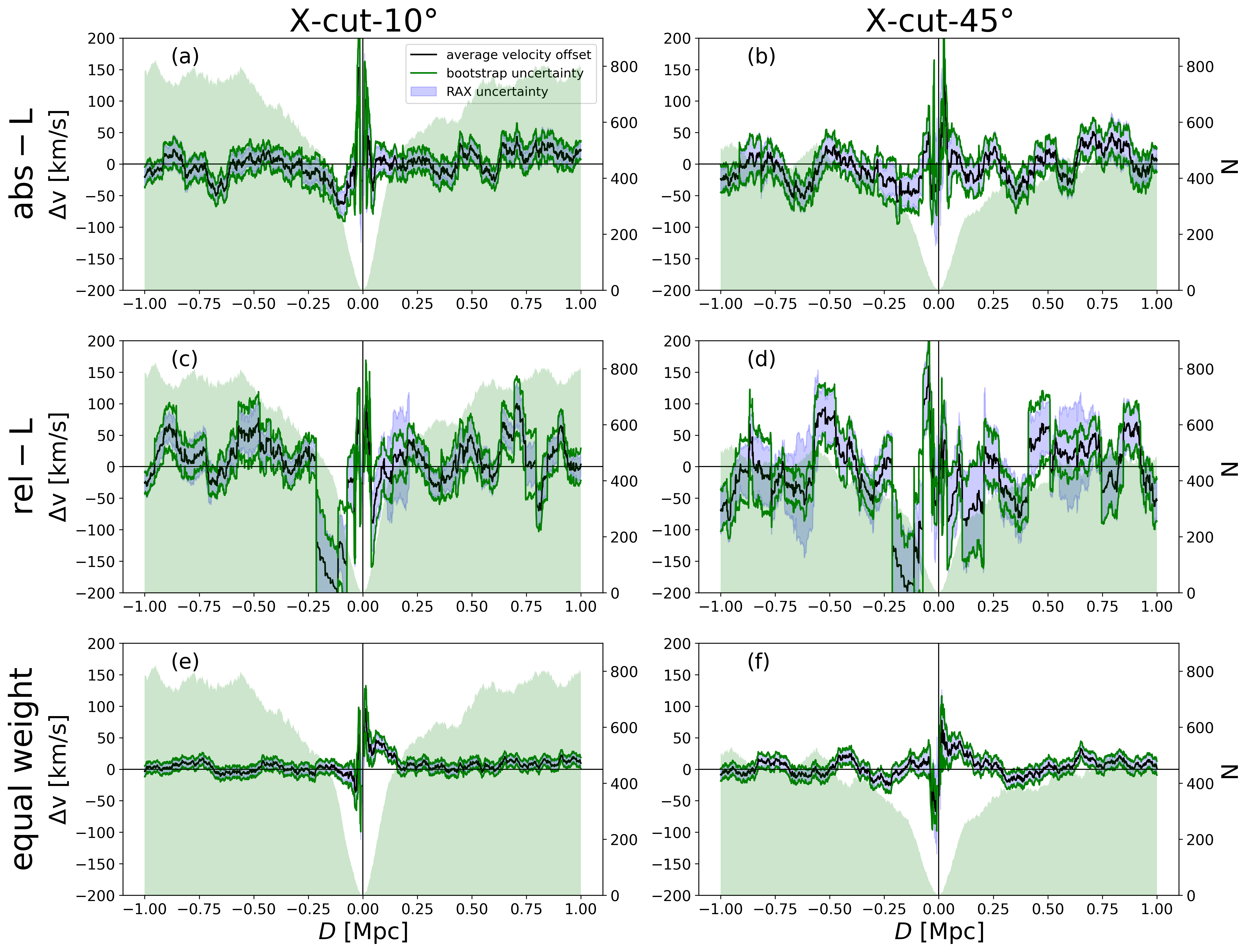}
\caption{The black lines show radial profile of average velocity offset within 1 Mpc as a function of projected distance, $D$. {\update The green lines show bootstrap uncertainties and the blue shades demonstrate RAX uncertainties.} The green histograms show the number of galaxies which contribute to the bin. The left column is based on X-cut-10° and the right column is based on X-cut-45°. The top, middle and bottom rows are abs-luminosity-weighted, relative-luminosity-weighted and equal-weighted mean velocity, respectively. If the correlation between the direction of spin vector of the central galaxy and the motion of its neighbour galaxies exists, then the average velocity offset would be positive on the right side and negative on the left side of each panel. However, we do not detect significant signal of coherence for the neighbouring galaxies within 1 Mpc scales. \label{fig:radialProfile1Mpc}}
\end{figure*}

We present average line-of-sight velocity of neighbouring galaxies as function of projected distance ${D}$ in Figure \ref{fig:radialProfile1Mpc}. If the coherence between the rotation of SAMI galaxies and the motion of neighbouring galaxies exists, we expect the average velocity {\update offset} to be positive on the right side of the profile and negative on the left. Figure \ref{fig:radialProfile1Mpc} shows the derived profiles based on X-cut-10° and X-cut-45° for abs-L, rel-L and equal weighting within 1 Mpc.

The extremely high velocity offset when the distance is smaller than 20 kpc is due to the small number of galaxies at this separation, leading to large uncertainty on the velocity. The number of galaxies contributing to the given bin is denoted by the green histograms in Figure \ref{fig:radialProfile1Mpc}. 

The profiles using the three different weighting scheme show broadly similar trends. However, the profile using the rel-L weighting fluctuates more than the abs-L weighting and equal weighting, because the mean velocity can be influenced by individual bright galaxies. The equal-weighted profile shows the least variance.

In Figure \ref{fig:radialProfile1Mpc}(a), the velocity offset at negative values of $D$ ranges from $\approx$ --50 to 20 km s$^{-1}$, and $\approx$ --20 to 30 km s$^{-1}$ for positive values of $D$. There are large outliers at $D$ = --0.15 Mpc in Figure \ref{fig:radialProfile1Mpc}(c), caused by a single bright galaxy ($M_r$ = --21.6 mag) nearby a faint SAMI galaxy ($M_r$ = --14.9 mag). Their rel-L weight counts nearly half the weight in that bin. The velocity offset between them is --256.5 km s$^{-1}$. These outliers are much smaller in abs-L weighting and disappear in equal weighting figures. Apart from these outliers, no significant signal of the coherence can be seen in Figure \ref{fig:radialProfile1Mpc}(c). The equal-weighted average velocity in Figure \ref{fig:radialProfile1Mpc}(e) is smaller than those in abs-L and rel-L. The average velocity profiles in X-cut-45° are similar to those in X-cut-10°.

\begin{figure*}
\centering
\includegraphics[width=0.9\textwidth]{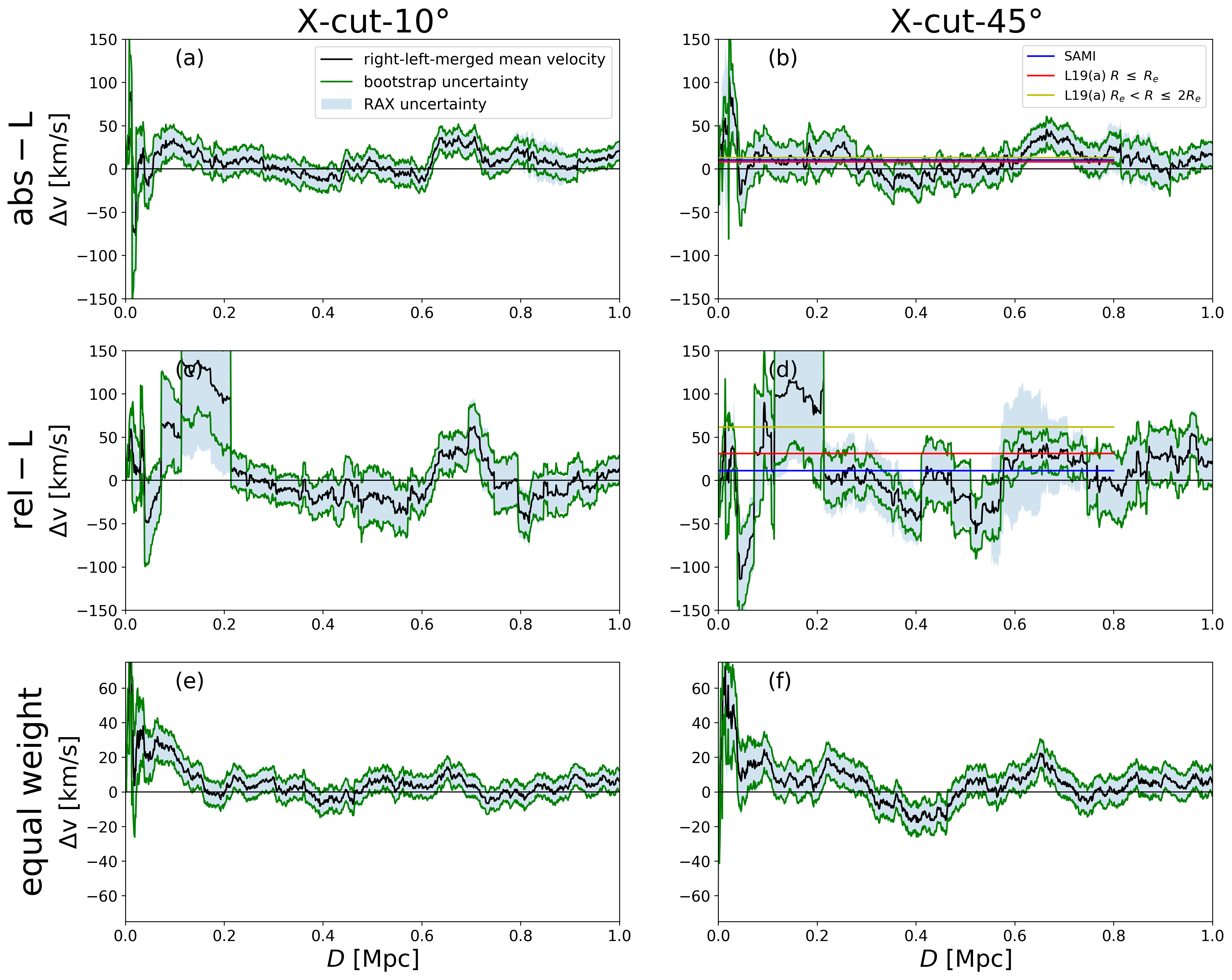}
\caption{Right-left-merged radial profile of average velocity offset within 1 Mpc. The left column is based on X-cut-10° and the right column is based on X-cut-45°. The top, middle and bottom rows are abs-luminosity-weighted, relative-luminosity-weighted and equal-weighted mean velocity, respectively. The horizontal lines in (b) and (d) demonstrate the cumulative right-left-merged mean velocities for the neighbour galaxies within 800 kpc scales, in which blue lines represent the results of SAMI galaxies, red and yellow lines represent the results in \citetalias{Lee:2019}. The uncertainties denoted by green lines are calculated by the bootstrapping method. The blue shades demonstrate RAX uncertainties. Note that the scale changes in (e) and (f).  \label{fig:RLMR1Mpc}}
\end{figure*}

Figure \ref{fig:RLMR1Mpc} presents the right-left-merged radial profile of velocity within 1 Mpc, calculated using Equation \ref{equ:R-L_radial}. Within a given interval, the right-left-merged average velocity is the average velocity from the positive $D$ directions minus the average velocity from the negative $D$ directions, and then normalized by the luminosity. If the coherence exists, the right-left-merged velocity would be positive. The uncertainties denoted by green lines are calculated by the bootstrapping method. The blue shading {\update demonstrates} RAX uncertainties. {\update Note that the bootstrap uncertainties are similar to RAX uncertainties in Figure \ref{fig:RLMR1Mpc} because we use small bins (100 kpc) when smoothing and plotting the profile. As a result the uncertainties dominated by random scatter rather than systematic effects.}

Figure \ref{fig:RLMR1Mpc}(a) shows the abs-L weighted average velocity in X-cut-10°. Even though there are some deviations away from zero at the level of up to $\sim2\sigma$, there is no obvious coherent signal. The velocity is consistent with zero at 200 < $D$ < 600 kpc. The outliers at $D$ = 150 kpc in rel-L (Figure \ref{fig:RLMR1Mpc}c and \ref{fig:RLMR1Mpc}d) are caused by bright neighbour galaxy nearby a single faint SAMI galaxy, as discussed above. The uncertainties when using rel-L is also larger than for the abs-L and equal weight cases. The dynamic range in weights is larger for the relative luminosity than the absolute luminosity. Therefore, the rel-L weighted average velocity is more likely driven by single galaxy and the bootstrap uncertainty is larger. Similarly, for the equal weighted panel in Figure \ref{fig:RLMR1Mpc}(e), we find the average velocity is 10--30 km s$^{-1}$ and is significant at 2$\sigma$ level at $D$ < 100 kpc. The average velocity is consistent with zero at 200 < $D$ < 1000 kpc. Table \ref{tab:compare} shows the cumulative right-left-merged mean velocity within a radius $D$, which is chosen to display the coherence signal. The signals of coherence for the neighbour galaxies within 1 Mpc are around the $1\sigma$ level for all three kinds of weighting cases.

\begin{center}
    \begin{table*}
        \centering
        \begin{tabular}{ccrrrr}
            \hline
             & & X-cut-10° & X-cut-45° & L19a ($R\leq R_e$) & L19a ($R_e< R\leq 2R_e$) \\
             Weighting & $D$ &  $\left \langle \Delta v \right  \rangle_{R-L}^{cumu}\pm\sigma_{RAX} $ & $\left \langle \Delta v \right  \rangle_{R-L}^{cumu}\pm\sigma_{RAX} $ & $\left \langle \Delta v \right  \rangle_{R-L}^{cumu}\pm\sigma_{BST} $ & $\left \langle \Delta v \right  \rangle_{R-L}^{cumu}\pm\sigma_{BST} $ \\
             & (Mpc) & (km s$^{-1}$) & (km s$^{-1}$) & (km s$^{-1}$) & (km s$^{-1}$)\\
            \hline
            abs-L   & 0.2 & $13.8\pm12.0\ (1.1\sigma)$ & $13.6\pm16.1\ (0.8\sigma)$ & $57.3\pm35.9\ (1.6\sigma)$ & $68.8 \pm37.3\ (1.8\sigma)$\\
            & 0.5 & $4.6\pm7.2\ (0.6\sigma)$ & $2.1\pm9.4\ (0.2\sigma)$ & $30.1\pm19.1\ (1.6\sigma)$ & $38.5\pm20.9\ (1.8\sigma)$\\
            & 0.8 & $8.6\pm5.8\ (1.5\sigma)$ & $10.4\pm8.6\ (1.2\sigma)$ & $8.3\pm12.8\ (0.6\sigma)$ & $13.1\pm14.2\ (0.9\sigma)$\\
            & 1 & $9.0\pm5.4\ (1.7\sigma)$ & $9.6\pm7.5\ (1.3\sigma)$ & - & -\\
            & 2 & $12.5\pm4.4\ (2.8\sigma)$ & $13.2\pm5.4\ (2.4\sigma)$ & - & -\\
            & 6 & $2.2\pm3.0\ (0.7\sigma)$ & $-2.6\pm3.6\ (-0.7\sigma)$ & $21.2\pm7.9\ (2.7\sigma)$ & -\\
            & 15 & $-4.2\pm2.7\ (-1.5\sigma)$ & $-7.1\pm4.2\ (-1.7\sigma)$ & - & -\\
            & 20 & $-4.3\pm2.9\ (-1.5\sigma)$ & $-6.6\pm4.0\ (-1.7\sigma)$ & - & -\\
            \hline
            rel-L   & 0.2 & $82.4\pm64.5\ (1.3\sigma)$ & $71.1\pm60.1\ (1.2\sigma)$ & $57.9\pm46.0\ (1.3\sigma)$ & $109.5\pm58.9\ (1.9\sigma)$\\
            & 0.5 & $21.0\pm24.8\ (0.8\sigma)$ & $17.3\pm29.2\ (0.6\sigma)$ & $25.2\pm23.4\ (1.1\sigma)$ & $61.1\pm27.8\ (2.2\sigma)$\\
            & 0.8 & $12.7\pm17.2\ (0.7\sigma)$ & $17.1\pm23.9\ (0.7\sigma)$ & $31.1\pm15.6\ (2.0\sigma)$ & $61.7\pm17.6\ (3.5\sigma)$\\
            & 1 & $11.1\pm14.4\ (0.8\sigma)$ & $18.6\pm21.1\ (0.9\sigma)$ & - & -\\
             & 2 & $10.5\pm10.5\ (1.0\sigma)$ & $27.9\pm15.9\ (1.8\sigma)$ & - & -\\
            & 6 & $2.7\pm6.2\ (0.4\sigma)$ & $-0.9\pm7.8\ (-0.1\sigma)$ & - & -\\
            & 15 & $-0.6\pm6.5\ (-0.1\sigma)$ & $-2.9\pm7.9\ (-0.4\sigma)$ & - & -\\
            & 20 & $-1.5\pm5.9\ (-0.2\sigma)$ & $-3.1\pm7.2\ (-0.4\sigma)$ & - & -\\

            \hline
             equal weight & 0.2 & $7.9\pm6.7\ (1.2\sigma)$ & $10.4\pm7.9\ (1.3\sigma)$ & - & -\\
             & 0.5 & $3.7\pm3.9\ (0.9\sigma)$ & $1.2\pm4.7\ (0.3\sigma)$ & - & -\\
             & 0.8 & $2.8\pm2.9\ (1.0\sigma)$ & $2.7\pm3.9\ (0.7\sigma)$ & - & -\\
             & 1 & $3.5\pm2.7\ (1.3\sigma)$ & $3.6\pm3.5\ (1.0\sigma)$ & - & -\\
            & 2 & $7.3\pm2.3\ (3.1\sigma)$ & $6.6\pm2.7\ (2.5\sigma)$ & - & -\\
            & 6 & $3.3\pm2.4\ (1.4\sigma)$ & $0.6\pm2.7\ (0.2\sigma)$ & - & -\\
            & 15 & $-1.7\pm2.4\ (-0.7\sigma)$ & $-3.9\pm3.7\ (-1.1\sigma)$ & - & -\\
             & 20 & $-2.8\pm2.5\ (-1.1\sigma)$ & $-4.0\pm3.6\ (-1.1\sigma)$ & - & -\\

            \hline
        \end{tabular}
        \caption{Cumulative right-left-merged mean velocity within a radius $D$, e.g. 0--0.8 Mpc. We also list the results in \citetalias{Lee:2019}. In L19a($R\leq R_e$), the angular momentum of CALIFA galaxies is measured at $R\leq R_e$. In L19a($R_e < R\leq 2R_e$), the angular momentum of CALIFA galaxies is measured at $R_e < R\leq 2R_e$.}
        \label{tab:compare}
    \end{table*}
\end{center}

\subsection{Large-scale results}

We also extend our study to a projected neighbour distance of 20 Mpc in order to investigate the velocity offset on larger scales. We calculate all three kinds of weightings, which qualitatively behave similarly to each other. Results for all three weightings are listed in Table \ref{tab:compare}, but to give an example of our results we only plot the results of the abs-L weighted analysis. The average velocity profile of relative luminosity weighting and equal weighting have similar trends to the profile of absolute luminosity weighting. 

\begin{figure*}
\centering
\includegraphics[width=0.9\textwidth]{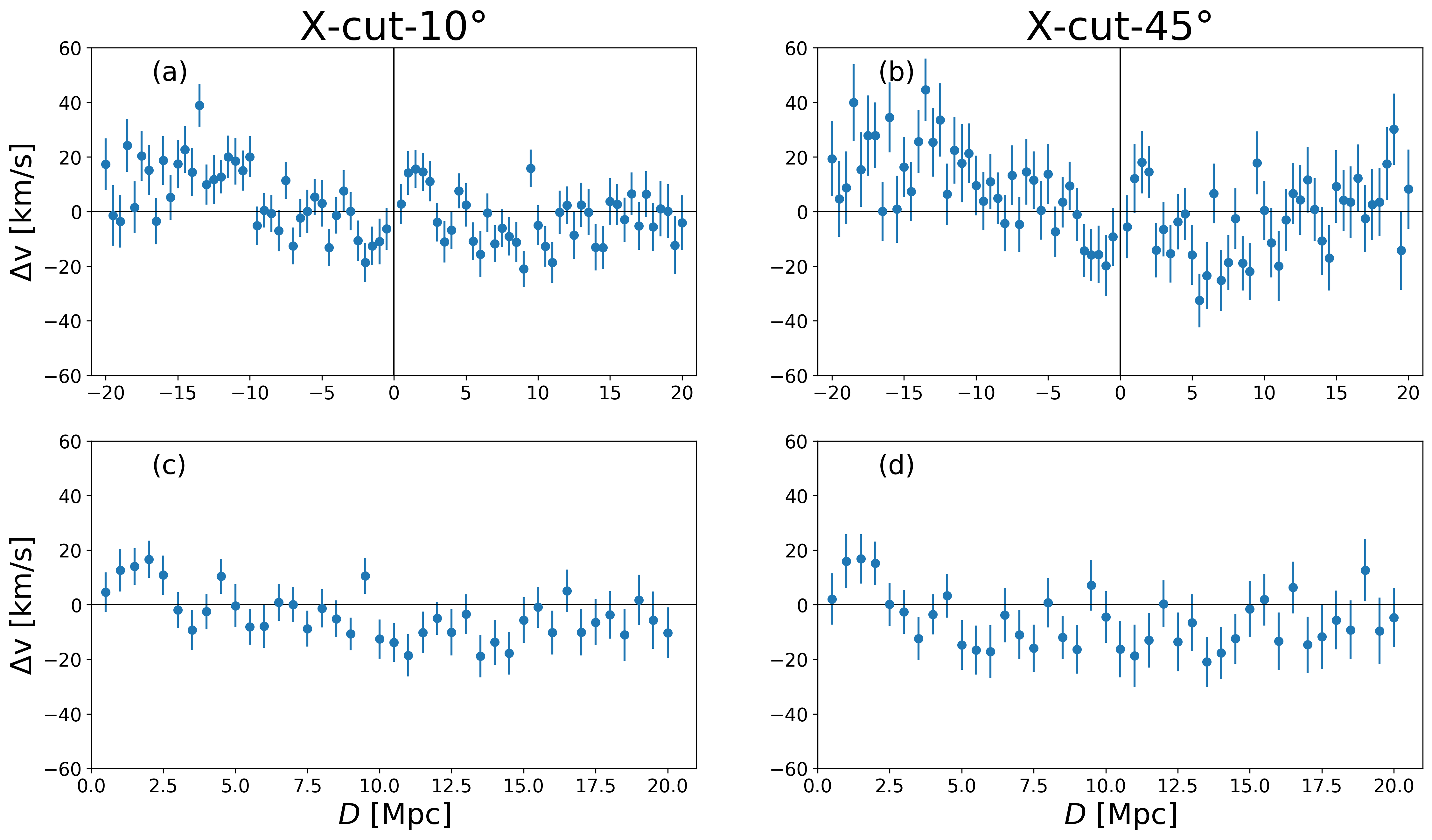}
\caption{We show 500-kpc-binned absolute-luminosity-weighted mean velocities within 20 Mpc scales in (a) and (b). Panel (c) and (d) show right-left-merged 500-kpc-binned mean velocities within 20 Mpc. The error bars represent $\sigma_{RAX}$.  \label{fig:RLMR20Mpc}}
\end{figure*}

When studying larger scale results we use 500-kpc-binned mean velocities, with no smoothing, so that each point is independent. The 500-kpc-binned mean velocities and right-left-merged 500-kpc-binned mean velocities are shown in Figure \ref{fig:RLMR20Mpc}. The 500-kpc-binned mean velocities show that there is a weak signal of coherence within 2 Mpc scales. The average velocities increase from 0--0.5 Mpc bin ($4.6\pm7.2$ km s$^{-1}$) to 1.5--2.0 Mpc bin ($16.6\pm6.8$ km s$^{-1}$) in the X-cut-10° case. Table \ref{tab:compare} also shows that the right-left-merged average velocity for the neighbour galaxies within 2 Mpc scales is $12.5 \pm 4.4$ km s$^{-1}$ in the X-cut-10° case, which is significant at the $\sim3\sigma$ level (RAX errors). 

The right-left-merged velocities are consistent with zero for the neighbouring galaxies in 3--8 Mpc distance. There is an “anti-correlation” between spin vector of SAMI galaxies and their neighbours motion for the neighbour galaxies outside 10 Mpc, i.e. the average velocity is positive on the left-side and negative on the right-side, as shown in Figure \ref{fig:RLMR20Mpc}(a) and (b). The right-left-merged average velocity is negative at 10 to 20\,Mpc, both in X-cut-10° and X-cut-45°, but their signals are also only significant at $\sim2\sigma$ level, as shown in Figure \ref{fig:RLMR20Mpc}(c) and (d).

\subsection{Results for galaxies in 6 different regions}

We divide the SAMI galaxies into 6 different regions on the sky, as shown in Figure \ref{fig:SpatialDist&z-mag}(a), in order to investigate the variation between different regions and see if the signal of coherence is stronger in any {\update region}. Table \ref{tab:1Mpc_ave} and Table \ref{tab:20Mpc_ave} list the right-left-merged average velocities for the different weights in each of these subsamples. We will focus on abs-L in the rest of this Section, but the results are qualitatively similar for the other weights. 

The average velocity offset varies between different regions. The standard errors on the mean of the right-left-merged velocities between 6 regions are 3.8 km s$^{-1}$ for X-cut-10° and 4.2 km s$^{-1}$ for X-cut-45° within 1 Mpc; 2.8 km s$^{-1}$ for X-cut-10° and 4.3 km s$^{-1}$ for X-cut-45° within 20 Mpc. The scatter between regions is similar to bootstrap uncertainty and RAX uncertainty of average velocity of whole sample, when considering neighbouring galaxies within 1 Mpc. The scatter between regions is similar with RAX uncertainty, but three times larger than bootstrap uncertainty when considering neighbouring galaxies within 20 Mpc, which also suggests that the bootstrap uncertainty of average velocity is an underestimate, particularly on large scales. 

In the abs-L case, Region 1 is a relative outlier at both 1 and 20 Mpc, which reflects the difference in large-scale structure of different regions. The average velocities in other regions are consistent with each other. Within 20 Mpc scales, the right-left-merged average velocities are negative in Region 1, 2, 4, 5 and 6. However, the velocity offsets are significant at 2$\sigma$ level in Region 1 and consistent with zero in others.

\begin{center}
\begin{table*}
    \centering
    \begin{tabular}{crrrrrr}
        \hline
         & \multicolumn{2}{c}{abs-L} & \multicolumn{2}{c}{rel-L} & \multicolumn{2}{c}{equal weight}\\
         \cmidrule(r){2-3} \cmidrule(r){4-5} \cmidrule(r){6-7}
         & X-cut-10° & X-cut-45°& X-cut-10° & X-cut-45°& X-cut-10° & X-cut-45°\\
         Region & $\left \langle \Delta v \right  \rangle_{R-L}^{cumu}\  $ & $\left \langle \Delta v \right  \rangle_{R-L}^{cumu}\  $ & $\left \langle \Delta v \right  \rangle_{R-L}^{cumu}\  $ & $\left \langle \Delta v \right  \rangle_{R-L}^{cumu}\  $& $\left \langle \Delta v \right  \rangle_{R-L}^{cumu}\  $ & $\left \langle \Delta v \right  \rangle_{R-L}^{cumu}\ $\\
         
         &(km s$^{-1}$) & (km s$^{-1}$)&(km s$^{-1}$) & (km s$^{-1}$)&(km s$^{-1}$) & (km s$^{-1}$)\\
        \hline
        1 & $-2.7\pm18.7$ & $-0.2\pm21.9$ & $-3.0\pm12.3$ & $0.4\pm14.6$ & $7.7\pm7.2$ & $-0.0\pm9.3$\\
        2 & $15.6\pm14.2$ & $-0.9\pm20.3$ & $-4.5\pm22.1$ & $-4.3\pm42.4$ & $-8.0\pm10.6$ & $-12.7\pm11.5$\\
        3 & $18.6\pm16.8$ & $24.3\pm20.5$ & $22.6\pm27.7$ & $49.7\pm64.4$ & $3.9\pm7.9$ & $4.8\pm11.9$\\
        4 & $18.8\pm14.2$ & $17.5\pm18.1$ & $20.6\pm40.4$ & $32.1\pm47.2$ & $7.4\pm7.5$ & $10.4\pm8.2$\\
        5 & $1.2\pm9.9$ & $2.8\pm11.7$ & $-2.2\pm12.0$ & $-6.1\pm14.2$ & $1.7\pm4.9$ & $3.6\pm6.0$\\
        6 & $6.6\pm14.2$ & $6.1\pm17.0$ & $17.2\pm20.8$ & $2.5\pm26.5$ & $1.9\pm7.1$ & $1.0\pm10.4$\\
        \hline
        mean & 9.7 & 8.3 & 8.4 & 12.4 & 2.4 & 1.2\\
        standard deviation & 9.3 & 10.3 & 12.9 & 23.0 & 5.7 & 7.7\\
        standard error on mean & 3.8 & 4.2 & 5.3 & 9.4 & 2.3 & 3.2\\

        \hline
    \end{tabular}
    \caption{The absolute-luminosity-weighted right-left-merged average velocity offset within 1 Mpc. The uncertainty of the right-left-merged average velocity is the RAX uncertainty ($\sigma_{RAX}$). The bottom of table shows the mean velocity of 6 different regions, the standard deviation and the standard error of the mean.}
    \label{tab:1Mpc_ave}
\end{table*}
\end{center}

\begin{center}
\begin{table*}
    \centering
    
    \begin{tabular}{crrrrrr}
        \hline
         & \multicolumn{2}{c}{abs-L} & \multicolumn{2}{c}{rel-L} & \multicolumn{2}{c}{equal weight}\\
         \cmidrule(r){2-3} \cmidrule(r){4-5} \cmidrule(r){6-7}
         & X-cut-10° & X-cut-45°& X-cut-10° & X-cut-45°& X-cut-10° & X-cut-45°\\
         Region & $\left \langle \Delta v \right  \rangle_{R-L}^{cumu}\  $ & $\left \langle \Delta v \right  \rangle_{R-L}^{cumu}\  $ & $\left \langle \Delta v \right  \rangle_{R-L}^{cumu}\  $ & $\left \langle \Delta v \right  \rangle_{R-L}^{cumu}\  $& $\left \langle \Delta v \right  \rangle_{R-L}^{cumu}\  $ & $\left \langle \Delta v \right  \rangle_{R-L}^{cumu}\ $\\
         
         &(km s$^{-1}$) & (km s$^{-1}$)&(km s$^{-1}$) & (km s$^{-1}$)&(km s$^{-1}$) & (km s$^{-1}$)\\
        \hline
        1 & $-14.2\pm7.0$ & $-22.0\pm8.4$ & $-16.2\pm10.3$ & $-13.9\pm12.3$ & $-10.8\pm6.1$ & $-15.6\pm7.3$\\
        2 & $-0.7\pm7.9$ & $2.8\pm9.9$ & $3.1\pm9.8$ & $0.7\pm12.2$ & $-1.7\pm6.9$ & $2.8\pm8.4$\\
        3 & $4.7\pm7.4$ & $4.8\pm10.1$ & $11.8\pm13.3$ & $7.7\pm18.4$ & $6.6\pm6.1$ & $5.7\pm8.2$\\
        4 & $-8.9\pm8.9$ & $-10.2\pm10.6$ & $0.8\pm12.2$ & $8.6\pm16.2$ & $-3.3\pm6.7$ & $-4.7\pm7.2$\\
        5 & $-0.8\pm6.1$ & $-4.3\pm7.5$ & $2.1\pm6.1$ & $-6.0\pm8.6$ & $0.7\pm5.7$ & $-1.8\pm6.8$\\
        6 & $-0.2\pm8.3$ & $4.5\pm9.0$ & $7.7\pm14.4$ & $7.3\pm17.0$ & $-2.4\pm6.8$ & $-0.7\pm6.9$\\
        \hline
        mean & -3.3 & -4.1 & 1.6 & 0.8 & -1.8 & -2.4\\
        standard deviation & 6.9 & 10.5 & 9.6 & 9.1 & 5.7 & 7.4\\
        standard error on mean & 2.8 & 4.3 & 3.9 & 3.7 & 2.3 & 3.0\\

        \hline
    \end{tabular}
    \caption{The absolute-luminosity-weighted right-left-merged average velocity offset within 20 Mpc. The uncertainty of the right-left-merged average velocity is the RAX uncertainty ($\sigma_{RAX}$). The bottom of table shows the mean velocity of 6 different regions, the standard deviation and the standard error of the mean.}
    \label{tab:20Mpc_ave}
\end{table*}
\end{center}

\subsection{Galaxy subsamples}
\label{sec:subsamples}

We also divide SAMI galaxies by galaxy mass, galaxy type, stellar spin parameter (quantified via $\lambda_{Re}$) and inclination ($i$) to find out if any subsample gives a more significant signal of coherence between the spin of SAMI galaxies and the motion of neighbours. Table \ref{tab:subsample_ave} shows the absolute-luminosity weighted right-left-merged mean velocity for the neighbouring galaxies within 1, 2, 6, 15 and 20 Mpc scales using X-cut-10°. 

We detect signals at the 2$\sigma$ level for low-mass (9 < log($M_*/M_\odot$)< 10.2) galaxies and high-mass (10.9 < log($M_*/M_\odot$)< 12) galaxies, but only 1.4$\sigma$ signal at intermediate masses (10.2 < log($M_*/M_\odot$)< 10.9) for neighbouring galaxies within 2 Mpc. It is not obvious what would cause this particular trend as a function of mass, although we note that the difference is not significant. The high-mass galaxies subsample gives $\sim2\sigma$ significance for the neighbouring galaxies within 1 Mpc and 2 Mpc. The average velocity of low-mass galaxies subsample is consistent with zero for the neighbours within 1 Mpc, but is significant at 2$\sigma$ level for the neighbours within 2 Mpc. However, the intermediate-mass galaxies subsample does not have any signal of coherence on any scale. 

The early-type galaxy subsample has a 3.5$\sigma$ signal for the neighbouring galaxies within 2 Mpc, which is the strongest detection in all of the subsamples. The signals of the late-type galaxies subsample are consistent with zero for the neighbouring galaxies within 1 and 2 Mpc.

For the $\lambda_{Re}$ case, {\update we split the sample into 3 equally sized subsamples.} The low-spin galaxies ($\lambda_{Re}$ < 0.409) have a stronger signal than the high-spin ($\lambda_{Re}$ > 0.586) and intermediate-spin (0.409 < $\lambda_{Re}$ < 0.586) for the neighbours within 1 Mpc and 2 Mpc scales.

We expect the edge-on galaxies (60° < $i$ < 90°) to have a stronger signal than the face-on galaxies (0° < $i$ < 30°), because the coherent motion of neighbours of face-on galaxies will be perpendicular to the line-of-sight, thus can not be measured by line-of-sight velocity offset. However, the edge-on galaxies subsample only has a 1.1$\sigma$ significance signal for neighbouring galaxies within 1 Mpc scales, which is not significantly larger than the face-on galaxies (0.6$\sigma$). For the neighbours within 2 Mpc scales, both the face-on galaxies subsample and the galaxies with inclination between 30° and 60° subsample have stronger signals (1.8$\sigma$ and 2.8$\sigma$, respectively) than the edge-on galaxies subsample (0.9$\sigma$).

When looking at large scales, we see the average velocity for the neighbouring galaxies within 20 Mpc of high-mass SAMI galaxies is negative and significant in 1.7$\sigma$, which indicates that the average velocity of neighbours between distance 10-20 Mpc is negative and significant at the same level with positive signal for neighbours within 2 Mpc. The negative signals may point to even the RAX errors underestimating the variance due to large-scale structure, so the significance of the positive signals might also be less than formally measured.  

The equal weighted results are a little different compared to the abs-L weighted results, as shown in Table \ref{tab:subsample_ave_unweighted}. The high-mass galaxies subsample does not have a signal for neighbours within 1 and 2 Mpc. The high-spin galaxies have slightly stronger signal than the low-spin galaxies, which is different to the abs-L weighted case. None of the differences between different subsamples is particularly convincing and the underlying mechanism of these differences is not clear.

{\update Finally, we apply our method to galaxy groups and galaxy pairs to see if the coherence signal is stronger for central galaxies and their satellites or galaxy pairs. 97.9\% of satellite galaxies are within 1 Mpc from their central galaxies and the distance between all galaxy pairs are smaller than 160 kpc. Since galaxies in a group or pair tend to have more interactions, we might expect to see a stronger coherence signal here. However, the right-left-merged average velocity offset for group galaxies is $-1.3 \pm 4.8$ km s$^{-1}$ and for paired galaxies is $9.3 \pm 6.1$ km s$^{-1}$, which do not show stronger signal than the other subsamples.}

To summarise, we do not find strong evidence that a specific subsample of galaxies shows a strong coherence signal. Our tentative results for high and low-mass galaxies (although not intermediate mass galaxies) and early-type galaxies would need to be confirmed with more observational data and studies using cosmological simulations.

\begin{table*}
	\centering
	
	\begin{tabular}{lrrrrr} 
		\hline
		 & 1 Mpc & 2 Mpc & 6 Mpc & 15 Mpc & 20 Mpc\\
		  & (km s$^{-1}$) &  (km s$^{-1}$)&  (km s$^{-1}$) & (km s$^{-1}$) & (km s$^{-1}$)\\
		\hline
        9 < log($M_*/M_\odot$)< 10.2&$-1.0\pm9.4\ (-0.1\sigma)$&$14.6\pm6.9\ (2.1\sigma)$&$3.0\pm5.0\ (0.6\sigma)$&$1.6\pm5.5\ (0.3\sigma)$&$-0.7\pm5.7\ (-0.1\sigma)$\\
        10.2 < log($M_*/M_\odot$)< 10.9&$7.4\pm7.4\ (1.0\sigma)$&$6.9\pm5.0\ (1.4\sigma)$&$-0.3\pm3.7\ (-0.1\sigma)$&$-5.4\pm4.3\ (-1.2\sigma)$&$-3.9\pm4.4\ (-0.9\sigma)$\\
        10.9 < log($M_*/M_\odot$)< 12&$29.9\pm13.2\ (2.3\sigma)$&$21.1\pm9.0\ (2.3\sigma)$&$6.2\pm4.8\ (1.3\sigma)$&$-9.1\pm5.8\ (-1.6\sigma)$&$-10.2\pm5.9\ (-1.7\sigma)$\\
		\hline
		early-type&$16.1\pm8.2\ (2.0\sigma)$&$22.2\pm6.4\ (3.5\sigma)$&$9.1\pm4.8\ (1.9\sigma)$&$-1.7\pm5.1\ (-0.3\sigma)$&$-3.0\pm5.1\ (-0.6\sigma)$\\
        late-type&$6.4\pm7.2\ (0.9\sigma)$&$5.5\pm5.2\ (1.1\sigma)$&$-3.4\pm4.3\ (-0.8\sigma)$&$-8.9\pm4.3\ (-2.1\sigma)$&$-8.5\pm4.3\ (-2.0\sigma)$\\
		
		\hline
		
		$\lambda_{Re}$ > 0.586&$9.8\pm11.0\ (0.9\sigma)$&$12.3\pm8.0\ (1.5\sigma)$&$1.6\pm4.5\ (0.4\sigma)$&$-5.5\pm4.8\ (-1.1\sigma)$&$-7.3\pm5.3\ (-1.4\sigma)$\\
        0.409 < $\lambda_{Re}$ < 0.586&$6.5\pm10.3\ (0.6\sigma)$&$10.8\pm8.9\ (1.2\sigma)$&$-0.1\pm6.1\ (-0.0\sigma)$&$-4.2\pm6.3\ (-0.7\sigma)$&$-2.4\pm6.3\ (-0.4\sigma)$\\
        $\lambda_{Re}$ < 0.409&$19.9\pm8.1\ (2.5\sigma)$&$18.0\pm7.4\ (2.4\sigma)$&$9.6\pm5.2\ (1.8\sigma)$&$-1.3\pm5.4\ (-0.2\sigma)$&$-2.0\pm5.9\ (-0.3\sigma)$\\
        \hline
        0° < $i$ < 30°&$8.6\pm13.6\ (0.6\sigma)$&$16.8\pm9.4\ (1.8\sigma)$&$3.9\pm5.9\ (0.7\sigma)$&$-5.6\pm5.2\ (-1.1\sigma)$&$-10.0\pm5.7\ (-1.7\sigma)$\\
        30° < $i$ < 60°&$7.6\pm6.4\ (1.2\sigma)$&$14.4\pm5.1\ (2.8\sigma)$&$5.7\pm3.8\ (1.5\sigma)$&$-1.3\pm3.7\ (-0.4\sigma)$&$0.4\pm3.9\ (0.1\sigma)$\\
        60° < $i$ < 90°&$14.7\pm13.1\ (1.1\sigma)$&$7.9\pm8.4\ (0.9\sigma)$&$-5.7\pm5.6\ (-1.0\sigma)$&$-8.4\pm6.0\ (-1.4\sigma)$&$-9.0\pm5.7\ (-1.6\sigma)$\\
		
		\hline
	\end{tabular}
	\caption{The absolute-luminosity weighted right-left-merged mean velocity for the neighbouring galaxies within 1, 2, 6, 15 and 20 Mpc scales of different subsamples.}
	\label{tab:subsample_ave}
\end{table*}

\begin{table*}
	\centering
	
	\begin{tabular}{lrrrrr} 
		\hline
		 & 1 Mpc & 2 Mpc & 6 Mpc & 15 Mpc & 20 Mpc\\
		  & (km s$^{-1}$) &  (km s$^{-1}$)&  (km s$^{-1}$) & (km s$^{-1}$) & (km s$^{-1}$)\\
		\hline
        9 < log($M_*/M_\odot$)< 10.2&$1.4\pm5.2\ (0.3\sigma)$&$8.8\pm4.2\ (2.1\sigma)$&$6.3\pm3.8\ (1.7\sigma)$&$1.7\pm4.4\ (0.4\sigma)$&$-0.8\pm4.7\ (-0.2\sigma)$\\
        10.2 < log($M_*/M_\odot$)< 10.9&$6.6\pm4.3\ (1.6\sigma)$&$4.4\pm3.3\ (1.4\sigma)$&$0.6\pm2.8\ (0.2\sigma)$&$-2.1\pm3.8\ (-0.6\sigma)$&$-2.1\pm4.0\ (-0.5\sigma)$\\
        10.9 < log($M_*/M_\odot$)< 12&$1.6\pm6.6\ (0.2\sigma)$&$7.0\pm6.6\ (1.1\sigma)$&$2.0\pm4.6\ (0.4\sigma)$&$-8.6\pm5.2\ (-1.7\sigma)$&$-9.9\pm5.5\ (-1.8\sigma)$\\
		\hline
		early-type&$4.7\pm4.8\ (1.0\sigma)$&$13.0\pm4.0\ (3.3\sigma)$&$7.6\pm4.3\ (1.8\sigma)$&$-0.1\pm4.6\ (-0.0\sigma)$&$-2.0\pm4.8\ (-0.4\sigma)$\\
        late-type&$2.0\pm4.8\ (0.4\sigma)$&$3.5\pm3.8\ (0.9\sigma)$&$0.1\pm3.5\ (0.0\sigma)$&$-5.2\pm3.6\ (-1.4\sigma)$&$-6.4\pm3.7\ (-1.7\sigma)$\\
		
		\hline
		
		$\lambda_{Re}$ > 0.586&$9.2\pm6.0\ (1.5\sigma)$&$9.1\pm4.7\ (2.0\sigma)$&$1.4\pm3.3\ (0.4\sigma)$&$-3.3\pm4.3\ (-0.8\sigma)$&$-5.7\pm4.9\ (-1.2\sigma)$\\
        0.409 < $\lambda_{Re}$ < 0.586&$2.2\pm5.7\ (0.4\sigma)$&$6.2\pm5.4\ (1.2\sigma)$&$0.2\pm5.2\ (0.0\sigma)$&$-2.8\pm5.8\ (-0.5\sigma)$&$-1.7\pm5.9\ (-0.3\sigma)$\\
        $\lambda_{Re}$ < 0.409&$4.8\pm4.5\ (1.1\sigma)$&$6.5\pm4.4\ (1.5\sigma)$&$9.7\pm4.2\ (2.3\sigma)$&$0.8\pm4.7\ (0.2\sigma)$&$-0.7\pm5.2\ (-0.1\sigma)$\\
        \hline
        0° < $i$ < 30°&$0.8\pm8.1\ (0.1\sigma)$&$11.1\pm6.2\ (1.8\sigma)$&$7.8\pm5.4\ (1.4\sigma)$&$-0.5\pm5.3\ (-0.1\sigma)$&$-5.3\pm5.8\ (-0.9\sigma)$\\
        30° < $i$ < 60°&$1.4\pm3.8\ (0.4\sigma)$&$6.7\pm3.5\ (1.9\sigma)$&$5.4\pm2.9\ (1.8\sigma)$&$0.9\pm3.0\ (0.3\sigma)$&$1.5\pm3.2\ (0.5\sigma)$\\
        60° < $i$ < 90°&$8.6\pm5.8\ (1.5\sigma)$&$5.7\pm4.7\ (1.2\sigma)$&$-3.2\pm4.4\ (-0.7\sigma)$&$-6.8\pm5.1\ (-1.3\sigma)$&$-8.2\pm5.0\ (-1.6\sigma)$\\
		
		\hline
	\end{tabular}
	\caption{The equal weighted right-left-merged mean velocity for the neighbouring galaxies within 1, 2, 6, 15 and 20 Mpc scales of different subsamples.}
	\label{tab:subsample_ave_unweighted}
\end{table*}

\section{Discussion} \label{sec:discussion}

\subsection{Comparison to the results of Lee et al.}

Table \ref{tab:compare} compares the coherence signals between our results and those in \citetalias{Lee:2019}. The limit of PA uncertainty in \citetalias{Lee:2019} is 45° and they used X-cut-45°. The $\sigma_{RAX}$ in our results are smaller, as the SAMI sample we use has a three times larger sample size compared to \citetalias{Lee:2019}. Also, the GAMA sample used to define environment reaches 2 magnitudes fainter than SDSS as well as having high completeness. Using abs-L weighting, the average velocities of our results are lower than those in \citetalias{Lee:2019}, except for average velocity for neighbours within 0.8 Mpc. In \citetalias{Lee:2019} the right-left-merged mean velocities are 57.3 $\pm$ 35.9 km s$^{-1}$ for the neighbour galaxies within 0.2 Mpc and 30.1 $\pm$ 19.1 km s$^{-1}$ for 0.5 Mpc, when they measured the PA of CALIFA galaxies using spaxels in centre regions. In our results, the velocities are 13.6 $\pm$ 16.1 km s$^{-1}$ for 0.2 Mpc and 2.1 $\pm$ 9.4 km s$^{-1}$ for 0.5 Mpc on X-cut-45°. Although the RAX uncertainty in our results are smaller than \citetalias{Lee:2019}, the significance of velocity offset in our results is formally consistent with zero. 

In the rel-L case, the right-left-merged mean velocity for neighbour galaxies within 0.2 Mpc for X-cut-45° is as large as $71.1 \pm 60.1$ km s$^{-1}$, which is similar with results in \citetalias{Lee:2019}. However, as we discussed in Section \ref{sec:results 1Mpc}, this signal is dominated by a single bright galaxy ($r$-band magnitude = --21.6 mag) nearby a faint SAMI galaxy ($r$-band magnitude = --14.9 mag). Therefore, the uncertainties in rel-L case are larger than those of abs-L case. The significance of signal in the rel-L case is also around the $1\sigma$ level. The equal weighted cases generally have smaller velocity offsets, and are formally consistent with zero. 

Thus, for scales less that 1 Mpc we do not find convincing evidence of a signal. While the measurements are not formally in statistical disagreement with those of \citetalias{Lee:2019}, we typically find a lower signal with a smaller uncertainty than \citetalias{Lee:2019}.

\citetalias{Lee:2019(b)} finds that abs-L weighted right-left-merged mean velocity is $21.2 \pm 7.9$ km s$^{-1}$ at $D$ $\leq$ 6.20 Mpc for central rotation and their 1-Mpc-binned right-left-merged mean velocity is positive up to 8 Mpc. However, we do not detect the same signal. Table \ref{tab:compare} shows that the abs-L weighted velocities are $2.2 \pm 3.0$ km s$^{-1}$ for X-cut-10° and $-2.6 \pm 3.6$ km s$^{-1}$ for X-cut-45° at $D\leq$ 6 Mpc. Comparing the most similar measurements at 6 Mpc (X-cut-45°) the difference between \citetalias{Lee:2019(b)} and our results is 2.7$\sigma$. Figure \ref{fig:RLMR20Mpc} also shows that the 500-kpc-binned right-left-merged mean velocity is positive up to 3 Mpc, but only significant in 1--3$\sigma$. The average velocity in 10.5--11.0 Mpc bin is $-18.6\pm7.8$ km s$^{-1}$, which is negative and significant at the same level as the average velocities for neighbours within 2 Mpc. As the signal within 2 Mpc has the same amplitude and significance as the negative signal in 10 Mpc, it is hard to distinguish the signal of coherence from the variance due to large-scale structure.

\subsection{Implication of subsamples}

We divide SAMI galaxies into 6 different regions on the sky and also divide SAMI galaxies by galaxy mass, galaxy type, $\lambda_{Re}$ and inclination to find out if any subsample has a stronger signal of coherence.

The mean velocity in region 1 at 20 Mpc is a relative outlier (see Table \ref{tab:20Mpc_ave}). This difference may be caused by large-scale structure near the SAMI galaxies. For instance, if the neighbours on the left side of the angular momentum of a SAMI galaxy are closer to us, while on the right side are far away, this will lead to positive mean velocity and vice versa. \citetalias{Lee:2019(b)} also found that 2 out of 6 regions in CALIFA galaxies have relatively strong coherence, while the other regions have ambiguous or no coherence signal. They suggested that large-scale coherence may be related to specific structure instead of being a universal property.

Although we expect the edge-on galaxies subsample to have the strongest signal among three inclination subsamples due to projection effects, it only has a 1.1$\sigma$ signal within 1 Mpc and is consistent with zero within 2 Mpc (see Table \ref{tab:subsample_ave}). However, the subsample of galaxies with inclination between 30° to 60° shows a 2.8$\sigma$ signal within 2 Mpc and the face-on galaxies subsample also has a 1.8$\sigma$ signal. These signals are more likely coming from coincidental scatter instead of arising from a particular physical property. We also notice that the average velocities of the edge-on and face-on galaxies subsamples for the neighbours within 20 Mpc are negative and significant at 1.6$\sigma$ and 1.7$\sigma$ level, which suggests that large-scale structure can also impact results.

The results of the three inclination subsamples imply that the 2--3$\sigma$ signal of low-mass, high-mass, early-type and low-spin galaxies subsamples are not enough to provide compelling evidence for a correlation between rotation of particular type of SAMI galaxies and the motion of neighbouring galaxies. However, if these signals are indeed real and confirmed by more observational results, there must be a physical explanation for this. We discuss the possible explanation in the following subsection.

\subsection{Comparison to previous studies}

{\update It is natural to think that the spin of galaxies could be influenced by the effects of galaxy-galaxy interactions over time. Using the IllustrisTNG simulation \citep{Nelson:2019}, \citet{Moon:2021} proposed that the alignment between the spin vector of a galaxy and the orbital motion of its neighbours results from the long-term interactions between paired galaxies. Their results showed that $\sim 10 \%$ more galaxies have their spin vector aligned with the orbital angular momentum vector of pair galaxies than a random isotropic sample. They found the alignment is stronger for low-mass central galaxies and galaxies in low-density environments.

Taking the findings of \citet{Moon:2021} we {\secupdate build a toy model} to compare to our findings in Section \ref{sec:results}. To make their results quantitatively comparable to ours, we map their results from the percentage of alignment to average right-left-merged velocity offset. We assume we have 1400 pairs of central galaxies and neighbouring galaxies, in which the spin vectors of 770 of the central galaxies are aligned with the orbital angular momentum vector of pair galaxies and for 630 pairs the vectors are anti-aligned (i.e., $10 \%$ more galaxies are aligned with their pair galaxies than random sample). The pairwise velocity dispersion or relative motion of galaxies varies with galaxy properties. On average, the three dimensional velocity dispersion between pair galaxies is around 500 km s$^{-1}$ \citep{Li:2006}. We now assume that this $500$ km s$^{-1}$ velocity is a reasonable estimation of the relative motion between galaxies and that all galaxies have this relative motion.  This is, no doubt, an oversimplification, but allows us to make a semi-quantitative comparison between our results and the simulations of \citet{Moon:2021}.  

We next consider the projection effects of galaxy inclinations and line-of-sight velocities by randomizing the viewing angle in our {\secupdate model} and calculating the equivalent of the equal-weighted velocity offset from Section \ref{sec:methods}. {\secupdate We build this toy model 1000 times by drawing the viewing angle from random distribution. Our model shows} that the average right-left-merged velocity offset is $20.3\pm6.6$ km s$^{-1}$ (where the uncertainty is standard deviation on the mean). 

The equal weighted average right-left-merged velocity offset for SAMI galaxies and their pair galaxies is $9.3\pm6.1$ km s$^{-1}$. Our result on the observational data in Section \ref{sec:results} is a little smaller than the above simulated results, but not significantly so, particularly given the simple assumptions used. We note, however, that our model ignores the effect of large-scale structure on the evolutionary history of our sample.
}

{\update {\secupdate It has recently become clear that the spin vector of galaxies is correlated with large-scale structure by both simulations and observations \citep[][Barsanti et al. submitted]{Codis:2012,Codis:2018,Dubois:2014,Welker:2020,BlueBird:2020,Kraljic:2020,Kraljic:2021,Tudorache:2022}}. Low-mass galaxies form in the neighbourhood of filaments, where the vorticity field is predominantly aligned with the filament \citep{Laigle:2015}. Then galaxies migrate towards filaments, accrete gas and merge with other low-mass galaxies along filaments. High-mass galaxies largely build their mass through mergers, so the spin axis tends to be dominated by angular momentum from the merger \citep{Codis:2018}.} {\update Barsanti et al.'s (submitted) finding that the flip of spin-alignment is most strongly correlated with the mass of galaxy's bulge also supports that mergers are the main drivers of the flip as the bulges of galaxies accrete mass through mergers.} \citet{Codis:2012} used the Horizon 4$\pi$ dark matter simulation and showed that dark matter flows towards filaments, forming low-mass dark matter haloes in the filaments. Galaxies in these halos tend to have their spin vector parallel to the filaments. Low-mass haloes merge with other haloes in the filaments and convert the orbital angular momentum into spin. Thus the high-mass haloes tend to spin perpendicular to the filaments. Hydrodynamical simulations \citep[e.g.][]{Dubois:2014} also showed that low-mass blue galaxies tends to align with filaments, while the alignment of high-mass red galaxies is more likely orthogonal to the filaments. Their results suggest that the tentative positive signal of coherence within 2 Mpc scales for the low-mass and high-mass galaxies might result from different physical mechanisms.

The spin alignment trends suggested by \citet{Dubois:2014} and \citet{Codis:2018} have been detected in a recent observational study by \citet{Welker:2020}. They investigated the kinematic spin-axis of the SAMI galaxies and their nearest cosmic filament in projection. They found that the spin-axis of low-mass SAMI galaxies tend to align with their nearest filament, whilst the spin-axis of high-mass galaxies tend to be orthogonal to their nearest filament. {\secupdate This pattern is consistent with that found in the Horizon-AGN simulation \citep{Dubois:2014,Codis:2018} and \textsc{SIMBA} simulation \citep{Kraljic:2020}.} This scenario suggests that low-mass galaxies are more likely to be aligned with cosmic web, and so they might be expected to show a stronger signal of coherence between galaxy spin and cosmic filament. We find some hints that low-mass galaxies have a stronger correlation between spin and the motion of neighbours (e.g. within 2 Mpc - see Table \ref{tab:subsample_ave} and \ref{tab:subsample_ave_unweighted}). These two results could be consistent with each other, but we caution that our measurement is different to that of \citet{Welker:2020}. We measure the how the spin of galaxies is related to the motion of neighbouring galaxies, whilst \citet{Welker:2020} measured how the spin of galaxies is related to the location of neighbouring galaxies. Detailed simulations of the two measurements would be needed to confirm that picture is consistent.

\citet{Xia:2020} used {\update the Millennium simulation \citep{Springel:2005}}, a dark-matter cosmological simulation to measure the angular momentum around filaments. They found dark matter rotates around the filament axis, implying that large-scale coherent motion exists in this simulation. Their result shows that the rotational velocity of dark matter haloes as a function of distance to the filament axis increases from 0 to 2 Mpc, peaking at around 55 km s$^{-1}$ and then decreasing. The trend of this velocity profile is similar with the trend in Figure \ref{fig:RLMR20Mpc}(c) within 3 Mpc scales. However, the average velocities in Figure \ref{fig:RLMR20Mpc}(c) are no more than 20 km s$^{-1}$. Note that our measurement is different to theirs: we measure the line-of-sight velocity offsets of neighbouring galaxies of each SAMI galaxy, whilst they measured the rotation of dark matter along the filaments in 3-dimensional space. Again, simulations of our specific measurements are needed to confirm any agreement.

An observational study using data from SDSS DR12 galaxy survey \citep{Wang:2021} also studies the spin of filaments. The spin of filaments seems to couple with the finding that the spin vector of low-mass galaxies tends to align with filaments. This can partially explain the tentative coherence signal we find of low-mass SAMI galaxies for the neighbouring galaxies within 2 Mpc.

Early-type galaxies, typically with higher mass and lower spin, have undergone on average $\sim$ 1 merger since $z\sim 1$ \citep{LopezSanjuan:2012}, which may influence the spin of galaxies. In addition, the early-type galaxies are more likely lived in denser environments \citep{Dressler:1980,Park:2007}, which exert stronger tidal force on galaxies and impact their spin. Both mergers and environment could potentially explain our tentative signal of coherence which is stronger in early-type galaxies than late-type galaxies, but we stress that a larger sample is required to confirm this finding and detailed simulations would demonstrate this in a quantitative way. Specifically, it would be valuable to make measurements of the coherence  of spin and cosmic flows in both large N-body simulations, looking at the effect on halos, and in recent large-volume hydrodynamical simulations such as EAGLE \citep{Schaye:2015} or IllustrisTNG \citep{Pillepich:2018}.

\section{Conclusion} \label{sec:Sum}

We use data from the SAMI Galaxy Survey to investigate the relationship between the direction of a galaxy's rotation and the average motion of its neighbours. We perform a similar analysis as \citetalias{Lee:2019}, but with a sample with three times as many galaxies and a neighbour population that reaches 2 magnitudes fainter than SDSS. When we calculate the average line-of-sight velocity offset between a SAMI galaxy and its neighbours, we weight them using three different methods: the first one considers the absolute luminosity of neighbour galaxies; the second considers the ratio of absolute luminosity between a neighbouring galaxy and a given SAMI galaxy; the third one is equal weighted. In addition, we calculate the velocity of different subsamples, including investigating subsets of the data in different regions of the sky and galaxies with different stellar mass, morphological type, $\lambda_{Re}$ and inclination. 

We find only very small velocity offsets that are significant up to $2\sigma$ level for the neighbouring galaxies within 1 Mpc, most of which are smaller than those found by \citetalias{Lee:2019}. When we extend the scope of neighbouring galaxies up to 20 Mpc, we find the average velocities increase up to 2 Mpc scales. The average velocity is $12.5\pm4.4$ km s$^{-1}$ (2.8$\sigma$) for the neighbouring galaxies within 2 Mpc (X-cut-10° case), but the average velocities for neighbours outside 3 Mpc are consistent with zero or negative. Some negative signals are also significant at 2$\sigma$ level. Therefore, we suggest that the deviations at the $\sim2\sigma$ level could be caused by differences in large-scale structure in the different observational regions from GAMA, and may not originate from a global relation between spin and neighbour motion.

We find modest evidence that low-mass, high-mass, early-type and low-spin galaxies subsamples have stronger signal of coherence for the neighbours within 2 Mpc. However, the results of different inclination subsamples and the large-scale results suggest that the 2--3$\sigma$ signals of some subsamples are not strong enough to distinguish coherence signals from coincidental scatter or variance of large-scale structure.

{\update We propose that the coherence signals in our results may result from the combined effect of galaxy interactions and large-scale structure. However,} simulation studies including N-body simulations and hydrodynamical simulations, as well as larger samples of IFS observations (e.g., the MaNGA survey \citealt{Bundy:2015}; the forthcoming Hector survey \citealt{Bryant:2020}) are needed to test our measurements and confirm these findings.

\section*{Acknowledgements}
The SAMI Galaxy Survey is based on observations made at the Anglo-Australian Telescope. The Sydney-AAO Multi-object Integral field spectrograph (SAMI) was developed jointly by the University of Sydney and the Australian Astronomical Observatory. The SAMI input catalogue is based on data taken from the Sloan Digital Sky Survey, the GAMA Survey and the VST ATLAS Survey. The SAMI Galaxy Survey is supported by the Australian Research Council Centre of Excellence for All Sky Astrophysics in 3 Dimensions (ASTRO 3D), through project number CE170100013, the Australian Research Council Centre of Excellence for All-sky Astrophysics (CAASTRO), through project number CE110001020, and other participating institutions. The SAMI Galaxy Survey website is http://sami-survey.org/ .

JvdS acknowledges support of an Australian Research Council Discovery Early Career Research Award (project number DE200100461) funded by the Australian Government. Sarah Brough acknowledges funding support from the Australian Research Council through a Future Fellowship (FT140101166). JBH is supported by an ARC Laureate Fellowship FL140100278. The SAMI instrument was funded by Bland-Hawthorn's former Federation Fellowship FF0776384, an ARC LIEF grant LE130100198 (PI Bland-Hawthorn) and funding from the Anglo-Australian Observatory. JJB acknowledges support of an Australian Research Council Future Fellowship (FT180100231). B.G. is the recipient of an Australian Research Council Future Fellowship (FT140101202).

\section*{DATA AVAILABILITY}
All observational data presented in this paper are available from Astronomical Optics’ Data Central service at https://datacentral.org.au/
as part of the SAMI Galaxy Survey Data Release 3.



\bibliographystyle{mnras}
\bibliography{galaxyRotation} 





\bsp	
\label{lastpage}
\end{document}